\documentclass[fleqn,usenatbib]{mnras}

\usepackage{enumitem,amssymb}
\usepackage{calc}
\newlist{todolist}{itemize}{2}
\setlist[todolist]{label=$\square$}
\usepackage{pifont}
\usepackage[T1]{fontenc}
\usepackage{newtxtext,newtxmath}
\usepackage{todonotes}
\usepackage{afterpage}
\usepackage{subcaption} 
\usepackage{float}
\usepackage{xspace}
\usepackage[normalem]{ulem}

\DeclareRobustCommand{\VAN}[3]{#2}
\let\VANthebibliography\thebibliography
\def\thebibliography{\DeclareRobustCommand{\VAN}[3]{##3}\VANthebibliography}

\newcommand{\borg}{\texttt{BORG}\xspace}

\newcommand{\Mpch}{\ensuremath{h^{-1}\,\text{Mpc}}}
\newcommand{\Gpch}{\ensuremath{h^{-1}\,\text{Gpc}}}
\newcommand{\hMpc}{\ensuremath{h\;\text{Mpc}^{-1}}}

\usepackage{graphicx}	
\usepackage{amsmath}	

\title[Quaia \borg{} Reconstruction]{Reconstructing the largest scales of the Universe with field-level inference applied to the \textit{Quaia} Quasar Catalogue}

\author[Andrews, Loureiro et al.]{
\parbox{2\columnwidth}{Adam Andrews,$^{1,2}$\thanks{E-mail: adam.andrews@inaf.it}
Arthur Loureiro,$^{3,4}$\thanks{E-mail: arthur.loureiro@fysik.su.se}
Jens Jasche,$^{3}$
Stuart McAlpine,$^{3}$
Guilhem Lavaux,$^{5}$
and Florent Leclercq$^{5}$
}\\
\\
$^{1}$ INAF/OAS Bologna, via Piero Gobetti 101, I-40129 Bologna, Italy\\
$^{2}$ INFN, Sezione di Bologna, via Irnerio 46, I-40126 Bologna, Italy\\
$^{3}$Oskar Klein Centre for Cosmoparticle Physics, Department of Physics, Stockholm University, Stockholm, SE-106 91, Sweden\\
$^{4}$ Astrophysics Group, Blackett Laboratory, Imperial College London, London SW7 2AZ, UK\\
$^{5}$ CNRS \& Sorbonne Université, UMR 7095, Institut d’Astrophysique de Paris, 98 bis boulevard Arago, F-75014 Paris, France}

\date{Accepted XXX. Received YYY; in original form ZZZ}

\pubyear{2026}

\begin{document}
\label{firstpage}
\pagerange{\pageref{firstpage}--\pageref{lastpage}}
\maketitle

\begin{abstract}
The recently released \textit{Quaia} quasar catalogue, with its broad redshift range and all-sky coverage, enables unprecedented three-dimensional reconstructions of matter across cosmic time. In this work, we apply the field-level inference algorithm \borg{} to the \textit{Quaia} catalogues to reconstruct the initial conditions and present-day matter distribution of the Universe. We employ a physics-based forward model of large-scale structure using Lagrangian perturbation theory, incorporating light-cone effects, redshift-space distortions, quasar bias, and survey selection effects.
This approach enables a detailed and physically motivated inference of the three-dimensional density field and initial conditions over the entire cosmic volume considered. We analyse both the $G<20.0$ (\textit{Quaia Clean}) and $G<20.5$ (\textit{Quaia Deep}) samples, where $G$ denotes the Gaia broad optical-band magnitude, imposing conservative sky cuts to ensure robustness against foreground contamination. The resulting reconstructions span a comoving volume of $(10h^{-1}~\mathrm{Gpc})^3$ with a maximum spatial resolution of $39.1\ h^{-1}\mathrm{Mpc}$, making this the largest field-level reconstruction of the observable Universe in terms of comoving volume to date. We validate our reconstructions through a range of internal and external consistency checks, including the cross-correlation of the inferred density fields with \textit{Planck} CMB lensing, where we detect a signal at $\sim4\sigma$ significance. Beyond delivering high-fidelity data products, including posterior maps of initial conditions, present-day dark matter, and velocity fields, this work establishes a framework for exploiting quasar surveys in field-level cosmology. 

\end{abstract}




\section{Introduction}
Mapping the large-scale structure of the Universe and the distribution of matter is a key step toward extracting fundamental physics from the largest cosmic scales. To accomplish this, various observational probes are commonly used, with quasars standing out as uniquely valuable tracers;  their extreme luminosities make them detectable up to redshifts $ z {\sim}7 $ \citep{2020ApJS..250....8L,2024ApJ...964...69S}, enabling studies on the largest observable scales. Moreover, by extracting information from their distribution, quasars are particularly useful for constraining primordial non-Gaussianity \citep{2014PhRvL.113v1301L,2014MNRAS.441..486K,2015JCAP...05..040H,2019JCAP...09..010C,2024JCAP...08..036C,2024JCAP...03..021K,2024arXiv241117623C,2025arXiv250214758C, 2025arXiv250420992F}, testing general relativity via relativistic effects \citep{2021MNRAS.501.1013Z,2024PhRvD.109h3540W}, and probing the turnover in the matter power spectrum $ P(k) $ \citep{2023MNRAS.524.2463B,2024arXiv241024134A}. Additionally, since quasars trace the largest scales, they can be cross-correlated with other cosmological probes, e.g. Cosmic Microwave Background (CMB) lensing \citep{2006PhR...429....1L,2020A&A...641A...8P}, to further constrain the growth of structure and the expansion history of the Universe \citep{2021JCAP...10..030G,2023Univ....9..302H,2023JCAP...11..043A,2024JCAP...03..021K,2024JCAP...06..012P}.

Despite being luminous tracers of the large-scale structure, quasar analysis is complicated by several issues. One key limitation is their low number density, which leads to higher levels of shot noise \citep{2018MNRAS.473.4773A,2024ApJ...964...69S}. Additionally, foreground contamination, e.g. Galactic dust and stellar sources, can introduce misidentifications and similar biases in quasar selection, introducing systematic effects that further complicate cosmological analyses \citep{2013MNRAS.435.1857L,2014MNRAS.444....2L}. Another major challenge is redshift determination: while spectroscopic redshifts provide precise distance measurements, the majority of quasars are identified through photometric methods, which introduce uncertainties that propagate into large-scale structure measurements \citep{2013MNRAS.435.1857L,2014MNRAS.444....2L,2023ApJ...944..107C}. Careful modelling is required to avoid systematic biases in cosmological parameter estimates.

The difficulties in analysing quasars illustrate a broader issue in large-scale structure studies; as datasets grow in size and precision, systematic uncertainties become the dominant challenge to overcome \citep{2013MNRAS.435.1857L,2014MNRAS.444....2L,2021MNRAS.506.3439R,2023ApJ...944..107C}. This is an aspect that will define the next generation of galaxy surveys, with examples of ongoing and upcoming large-scale structure surveys include the \textit{Euclid Space Telescope} \citep{2024arXiv240513491E}, the \textit{Nancy Grace Roman Space Telescope} \citep{2019BAAS...51c.341D}, SPHEREx \citep{2018arXiv180505489D}, \textit{China Space Station Telescope} (CSST, \citealt{2025arXiv250115023G}), and the \textit{Vera C. Rubin Observatory} \citep{2018arXiv180901669T}. These surveys will dramatically improve our ability to map the cosmic web by observing billions of galaxies across the Universe. Although current analyses are often limited by statistical noise \citep{2018AJ....155....1G}, the increase in volume and number density of upcoming surveys will shift the limiting factor from statistical to systematic uncertainties, including foreground contamination, selection effects, and redshift errors. Quasars are a valuable probe within these surveys, complementing galaxy samples by tracing the large-scale structure at higher redshifts \citep{2018MNRAS.473.4773A,2024ApJ...964...69S,2025arXiv250314745D}.

In recent years, Field-Level Inference (FLI) has emerged as a novel approach for extracting cosmological information from large-scale structure surveys, offering a direct alternative to summary statistics such as power spectra and bispectra. These traditional analysis methods are based on suboptimal data compression, failing to fully exploit high-dimensional information in large-scale surveys \citep{2024arXiv240502252B}. Meanwhile, by directly forward-modelling the observed galaxy field, the FLI approach can integrate physical models of structure formation, survey geometry, and foreground templates \citep{2008MNRAS.389..497K,2010MNRAS.406...60J,2010MNRAS.407...29J,2012MNRAS.425.1042J,2013MNRAS.432..894J,2015JCAP...01..036J,2015JCAP...03..047L,2017A&A...606A..37J,2019JCAP...10..035H,2019arXiv190906396L,2019A&A...624A.115P,2019A&A...625A..64J,2023JCAP...10..069S,2022JCAP...08..007B,2025JCAP...11..082C,2025JCAP...04..089S,2025JCAP...11..055S,2025MNRAS.544..960H}. In this way, FLI accounts for the underlying physical processes that shape the observed matter distribution, and thus can access information that is otherwise lost in data compression \citep{2021MNRAS.506L..85L,2024PhRvL.133v1006N,2024arXiv240502252B}. Thus, this novel approach enables a more comprehensive and robust analysis of the data, taking full advantage of the information contained in the data. 

FLI has been applied to a wide range of problems, including the reconstruction of cosmic initial conditions \citep{2010MNRAS.406...60J,2013MNRAS.432..894J,2015JCAP...01..036J,2016MNRAS.455.3169L}, constraints on primordial non-Gaussianity \citep{2023MNRAS.520.5746A,2024arXiv241211945A,2025JCAP...03..016S}, modified gravity \citep{2025MNRAS.544.3634S}, baryon-acoustic oscillations \citep{2022JCAP...08..007B, 2025JCAP...11..066B,2025arXiv250513588B}, and the study of dark energy \citep{2019A&A...621A..69R}. FLI has also been used in weak lensing \citep{2021MNRAS.502.3035P,2022MNRAS.509.3194P,2022arXiv221012280F,2022MNRAS.512...73F,2022MNRAS.516.4111B,2023OJAp....6E...6L,2024PhRvD.110b3524B,2024MNRAS.527L.162B}, Lyman-$\alpha$ forest analysis \citep{2019A&A...630A.151P,2020A&A...642A.139P,2025arXiv250700284D}, HI intensity mapping \citep{2023PhRvD.108h3528O}, on angular sky maps \citep{2023OJAp....6E..31S,2025arXiv251026691C,2025PhRvD.111h3542B}, to test new physics \citep{2020PhRvD.102j4060D,2021PhRvD.103b3523B,2021PhRvD.104j3516B,2022PhRvD.106j3526B,2023arXiv230410301K}, and peculiar velocity reconstructions \citep{2022MNRAS.517.4529B,2023MNRAS.518.4191P,2025MNRAS.tmp.1852S,2025arXiv250909665S}. Furthermore, advances in Bayesian posterior sampling have enabled FLI to be used for posterior resimulations \citep{2022MNRAS.512.5823M,2024A&A...691A.348W,2025MNRAS.540..716M} and in combination with machine learning approaches to accelerate inference \citep{2024MNRAS.535.1258D,2025arXiv250213243D}. The methodology has also been developed for photometric surveys \citep{2012MNRAS.425.1042J,2023arXiv230103581T}. There is also an ongoing investigation into the theoretical foundations, information content, statistical consistency, and computational scalability of field-level inference in large-scale structure analyses \citep{2021JCAP...03..058N,2023JCAP...07..063K,2025arXiv251105484R,2025JCAP...09..056S,2025arXiv250909673A,2025arXiv250705378S,2025arXiv250420130S}.

Beyond theoretical applications, FLI has been successfully implemented in real observational data, demonstrating its viability \citep{2015JCAP...01..036J,2016MNRAS.455.3169L,2019A&A...625A..64J,2019arXiv190906396L}. Notable examples include the application of Bayesian forward modelling techniques to reconstruct the cosmic web, as well as the inference of cosmic initial conditions, from spectroscopic galaxy redshift surveys such as 2M++ and BOSS \citep{2016MNRAS.455.3169L,2019arXiv190906396L,2019A&A...625A..64J}. Field-level inference has also enabled posterior resimulations, allowing for detailed studies, including local galaxy formation \citep{2022MNRAS.512.5823M,2022MNRAS.509.1432S,2025arXiv250314732G}, intrinsic alignment effects \citep{2022JCAP...08..003T,2023arXiv230404785P}, and even generating a digital twin of our local Universe \citep{2016MNRAS.455.3169L,2022MNRAS.511L..45D,2022MNRAS.516.3592H,2024MNRAS.527.1244S,2024MNRAS.531.2213S,2025MNRAS.540..716M}. There are also works that employ hierarchical and cosmic-web biasing schemes to address the galaxy bias problem when modeling galaxy formation across a wide redshift range \citep{2024JCAP...07..083C,2024JCAP...07..001F,2025arXiv250603969R,2025arXiv250917696A,2025arXiv250915890S}. As all of these methods continue to mature, they provide a pathway toward fully exploiting high-dimensional cosmological data, maximising the information content available by large-scale structure studies by galaxy surveys.
 
A specific quasar dataset designed for large-scale structure studies is the \textit{Quaia} catalogue presented in \cite{2024ApJ...964...69S}, which consists of quasars selected from \textit{Gaia} \citep{2023A&A...674A..31D} and unWISE \citep{2014AJ....147..108L,2019PASP..131l4504M}. \textit{Quaia} covers a wide redshift range, extending up to $ z \approx 4 $, and provides a large sky coverage of $ f_{\rm sky} \approx 0.71 $, making it a valuable dataset for cosmological analyses \citep{2023JCAP...11..043A,2024arXiv241024134A}. Due to its high completeness and well-characterised selection function, \textit{Quaia} has been used to study the growth of structure and test models of gravity \citep{2024MNRAS.527.8497M}, measure cosmological dipole \citep{2021ApJ...908L..51S,2022ApJ...937L..31S,2024MNRAS.527.8497M,2024JCAP...11..067A} constraining primordial non-Gaussianities \citep{2025arXiv250420992F}, and particularly robust analysis when cross-correlated with CMB lensing \citep{2023JCAP...11..043A,2024JCAP...06..012P}. Another widely used quasar dataset is SDSS-IV/eBOSS \citep{2020ApJS..250....8L}, which has played a key role in constraining local primordial non-Gaussianity through power spectrum and bispectrum estimators \citep{2019JCAP...09..010C,2024JCAP...08..036C,2025arXiv250214758C} and turnover of the 3D power spectrum, $P(k)$ \citep{2023MNRAS.524.2463B}.

In this study, we apply \textit{Bayesian Origin Reconstruction of Galaxies} (\borg{}, \citealt{2013MNRAS.432..894J,2015JCAP...01..036J,2016MNRAS.455.3169L,2019A&A...625A..64J,2019arXiv190906396L}), a field-level inference framework, to the \textit{Quaia} quasar catalogue. By forward-modelling the physical formation of structure and accounting for survey selection effects and foreground contamination, we infer the three-dimensional matter density field, initial conditions, and velocity fields traced by quasars. This work represents the first application of field-level inference to a quasar dataset and the largest reconstruction by survey volume to date. The unprecedented volume of \textit{Quaia} enables exploration of ultra-large scales, while requiring careful treatment of observational systematics.

This paper is organised as follows. Section \ref{sec:data} describes the \textit{Quaia} dataset, including completeness masks, radial selection functions, and foreground templates. Section \ref{sec:method} outlines the \borg{} methodology and analysis pipeline. In Section \ref{sec:results}, we present the reconstructed fields, while Section \ref{sec:validation} assesses their robustness through comparisons with theoretical expectations and posterior predictive tests. Section \ref{sec:cmb_cross} presents the detailed CMB lensing cross-correlation analysis \citep{2020A&A...641A...8P}. Finally, Section \ref{sec:conclusions} summarises our results and discusses future directions. The appendices provide additional results (\ref{Appx:AddResults}), details on systematic mitigation (\ref{Appx:SystContamination}) and methodology (\ref{Appx:TechDets}), convergence diagnostics (\ref{Appx:convergence}), further checks of the velocity fields (\ref{Appx:velocity_dipole}), and CMB lensing signal modelling (\ref{Appx:cmb_signal_modelling}).

\begin{figure*}
	\centering
        \includegraphics[trim={3.25cm 2cm 0.5cm 2cm},clip,width=2\columnwidth]{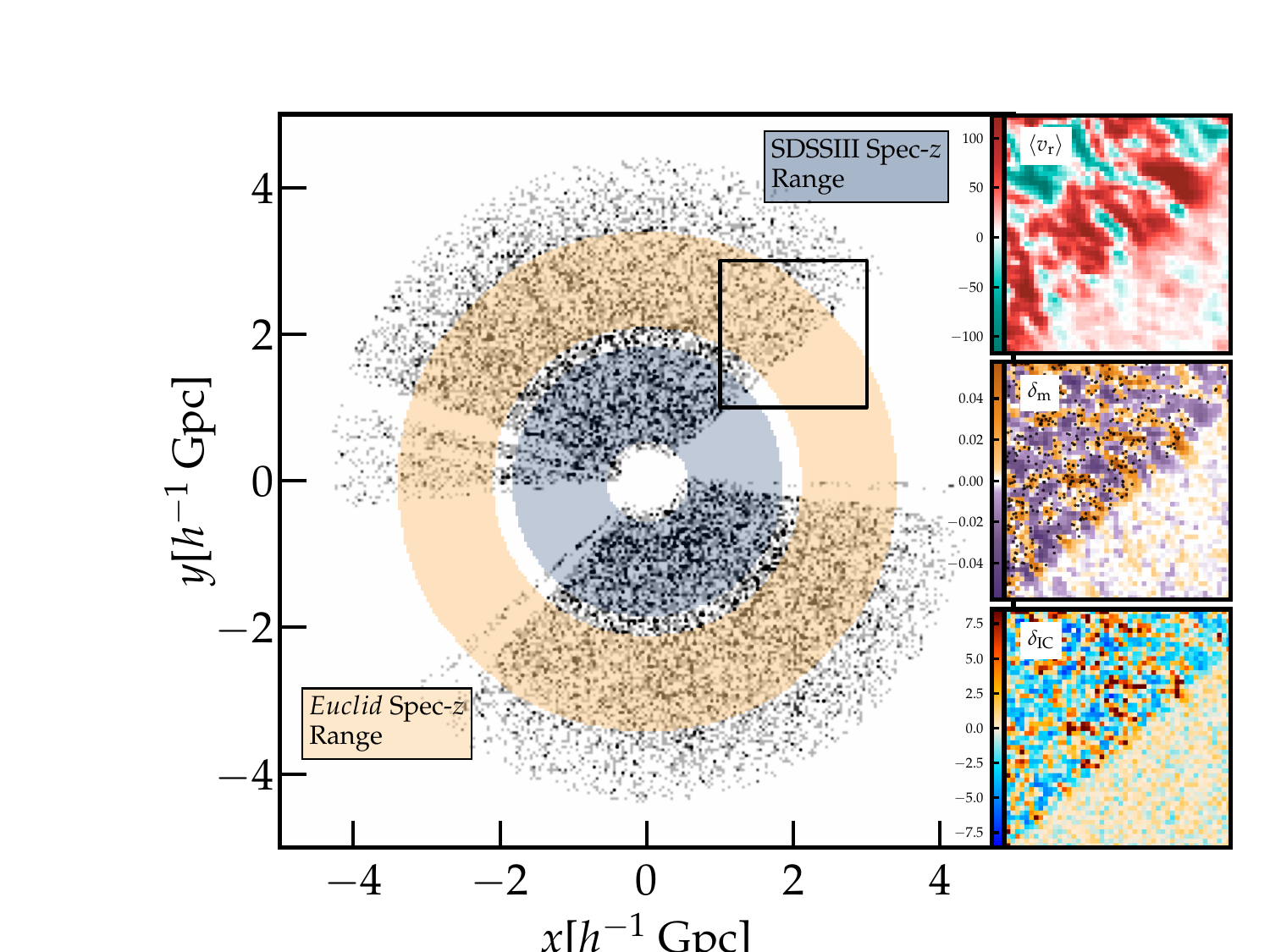}
	\caption{Distribution of quasars and inferred fields from the \textit{Quaia Deep Cut} quasar catalog, which provide an overview of the outcomes of the \borg{} analysis. The main panel shows the survey volume coverage, comparing \textit{Quaia} (black dots) with SDSS-III (gray range) and the \textit{Euclid} spectroscopic range (yellow range). The panels on the right-hand side shows the corresponding inferred fields of the black inset; the bottom-right panel shows the mean inferred initial conditions ($\delta_{\mathrm{IC}}$), centre-right the dark matter density field ($\delta_{\rm{m}}$), and upper-right the radial velocity field ($v_{\rm r}$). The imprint of the survey edge in the panels highlights \borg{}’s ability to account for systematics across the ensemble of the MCMC samples. In the dark matter density panel, we overlay the quasars, illustrating the match between input data and inferred structures, which demonstrates the ability of the \borg{} to recover the large-scale structure from sparsely distributed quasars. The thickness of the slice shown here is approximately $39\, \Mpch$; the unit for the radial velocity field is $\rm{km\,s^{-1}}$.
}
	\label{fig:first_plot}
\end{figure*}
\section{\textit{Quaia}, a Gaia-unWISE Quasar Catalogue}\label{sec:data}

In this study, we utilise the all-sky \textit{Quaia} \textit{Gaia}-unWISE quasar catalogue, as presented by \cite{2024ApJ...964...69S} (hereafter SF24). This catalogue is derived from \textit{Gaia} Data Release 3 quasar candidates, which identify more than six million potential Quasi-Stellar Objects (QSOs). Because \textit{Gaia} provides low-resolution spectra, SF24 enhanced its redshift estimates by combining \textit{Gaia} DR3 data with mid-infrared data from the unWISE reprocessing of the Wide-field Infrared Survey Explorer (WISE) observations, thereby improving both sample purity of the sample and redshift accuracy (see Section \ref{sect:SPZ} for more details).

 SF24 provides two versions of the \textit{Gaia}-unWISE Quasar Catalogue, which we consider in our analysis: \textit{Quaia Deep} and \textit{Quaia Clean}. \textit{Quaia Deep} comprises sources brighter than $G < 20.5$, covering the entire sky with high homogeneity and containing $1\,295\,502$ sources with a median redshift of $z_{\rm med} = 1.48$ \citep[used also in][]{2023JCAP...11..043A}. In contrast, \textit{Quaia Clean}, which adopts a stricter magnitude limit of $G < 20.0$, includes about $755\,850$ sources with a median redshift of $z_{\rm med} = 1.45$. Both catalogues span a sky area of $29\,154.54 \,\deg^2$ (corresponding to $f_{\rm sky} \approx 0.71$) and a redshift range of $0.084 < z \lesssim 4.0 $. In Figure \ref{fig:first_plot}, we show a slice of the \textit{Quaia Deep Cut} in comparison with the redshift range of SDSS-III and Euclid-Spec, together with zoomed panels into the main inferred data products presented in Section \ref{sec:results}.

\subsection{Sample selection}
We consider two versions of the \textit{Quaia} catalogue in our analysis. The first is the \textit{Quaia Clean} sample, defined by a magnitude cut of $G < 20.0$, and accompanied by an angular completeness mask constructed by SF24. This mask was derived using a Gaussian process fit that incorporates several systematic map templates, accounting for observational effects such as \textit{Gaia} DR3 stellar density, the unWISE source distribution, source density variations around the Large and Small Magellanic Clouds (LMC and SMC), and the scan patterns of both \textit{Gaia} and WISE.

\begin{figure*}
    \centering
    \begin{subfigure}{0.48\textwidth}
        \centering
        \includegraphics[width=\textwidth]{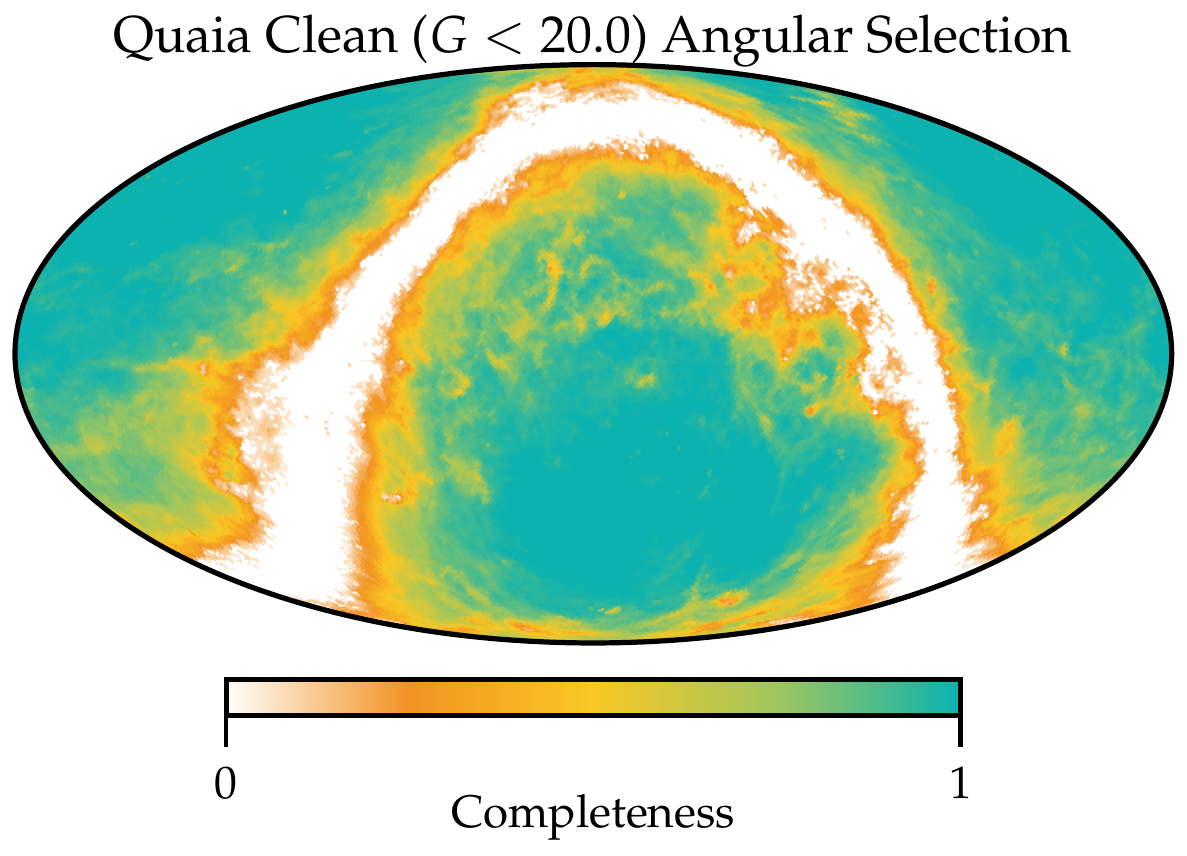}
        \caption{Angular Selection Function for the \textit{Quaia Clean} Sample}
        \label{fig:QuaiaCleanCompletness}
    \end{subfigure}
    \hfill
    \begin{subfigure}{0.48\textwidth}
        \centering
        \includegraphics[width=\textwidth]{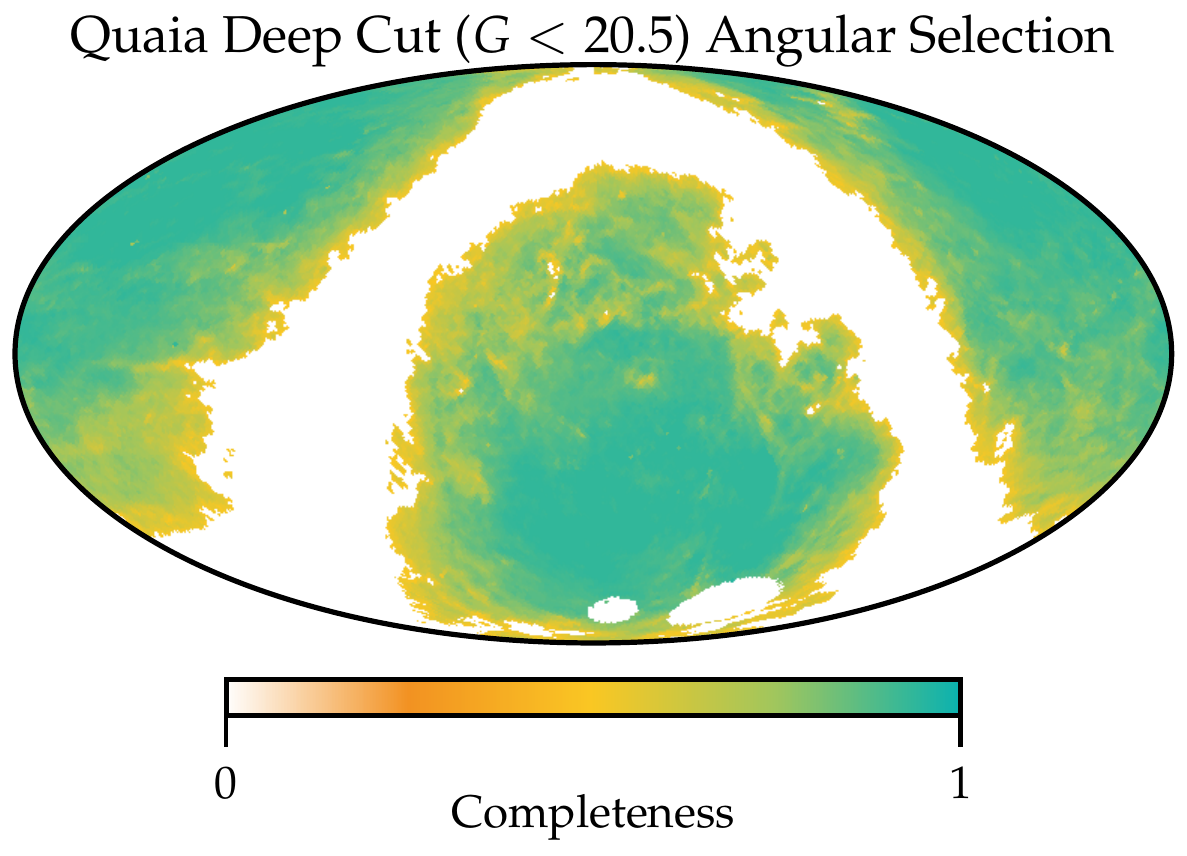}
        \caption{Angular Selection Function for the \textit{Quaia Deep Cut} Sample}
        \label{fig:QuaiaDeepCutCompletness}
    \end{subfigure}

    \caption{Angular selection function, or completeness, for the \textit{(a)} \textit{Quaia Clean} ($G<20.0$) sample, as estimated by SF24, and the \textit{(b)} \textit{Quaia Deep Cut} ($G<20.5$) sample. Both maps are shown in celestial coordinates. Note that for the \textit{Quaia Deep Cut} we removed the Large and Small Magellanic Clouds and applied a completeness cut of $0.5$. See text for details.}
    \label{fig:20_5_mask}
\end{figure*}

The second sample, which we refer to as the \textit{Quaia Deep Cut}, extends the magnitude limit to $G < 20.5$ while applying additional angular selection criteria. Specifically, we impose a completeness threshold of $> 0.5$ to exclude regions where the original completeness is too low, and we mask out areas near the LMC and SMC, which show significant contamination in preliminary analyses (see Appendix~\ref{Appx:SystContamination} for details). The resulting completeness masks for both samples are shown in Figures~\ref{fig:QuaiaCleanCompletness} and~\ref{fig:QuaiaDeepCutCompletness}, respectively.

\subsection{Data preparation}
\begin{figure}
	\centering
        \includegraphics[width=\columnwidth]{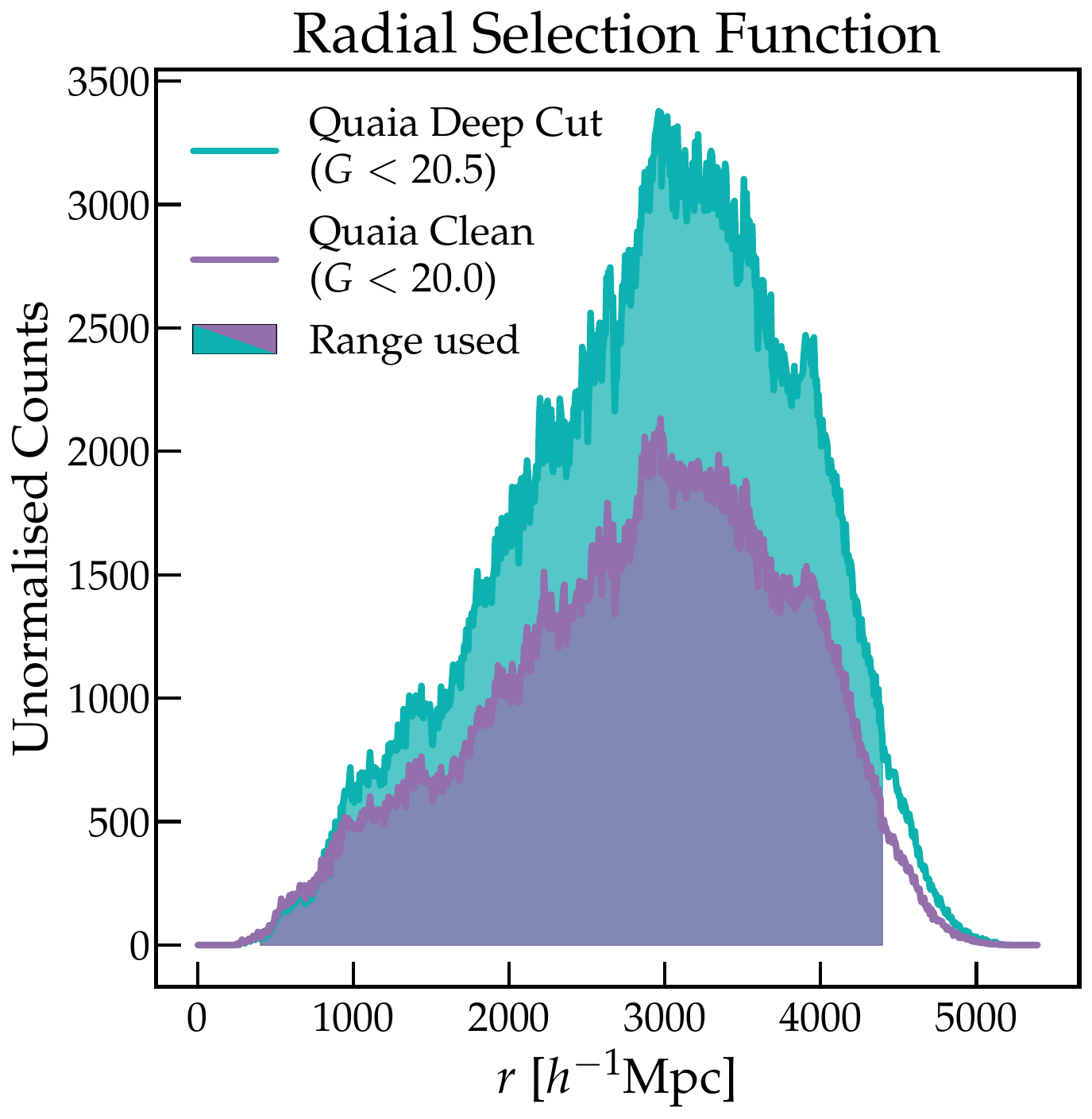}
	\caption{Radial selection function for the \textit{Quaia Clean} ($G<20.0$) sample (\textit{purple}) and the \textit{Quaia Deep Cut} ($G<20.5$)  sample (\textit{teal}). The shaded region outlines the range used in the sub-catalogues in our analysis.}
	\label{fig:20_5_rsf}
\end{figure}

For all samples considered, we partition the data into eight distinct radial bins to probe different cosmic epochs while maintaining sufficient statistical power in each subcatalogue. The radial bins, configured to maintain approximately equal comoving volumes, span radial distances from $400$ to $4400\,h^{-1} \text{Mpc}$, with each bin having a radial width of $\chi_{\rm bin} = 500\,h^{-1} \text{Mpc}$. During inference, each radial bin is treated as an independent catalogue: within \borg{}, all galaxies (or quasars) in the same bin share the same bias parameters, which are sampled sequentially using a slice sampler while keeping the other bins and parameters fixed. Importantly, all catalogues share the same underlying density field, which ensures a consistent reconstruction across the full radial range. The full radial selection functions for both \textit{Quaia Clean} and \textit{Quaia Deep Cut} are shown in Figure~\ref{fig:20_5_rsf}. 

Additionally, assuming that the quasars are on average stationary with respect to the CMB, we transform all quasar redshifts into the rest frame of the CMB to remove the imprint of our own motion (the Sun’s peculiar velocity) from the observed redshifts. While this assumption is still under discussion for some quasar samples \citep{2021ApJ...908L..51S,2022ApJ...937L..31S}, it appears to be reasonable for the \textit{Quaia} catalogues used here \citep{2024MNRAS.527.8497M}.

In the heliocentric (or geocentric) frame, that motion induces a Doppler dipole: quasars in the direction we are moving towards appear slightly blueshifted, and those opposite to us appear slightly redshifted. If left uncorrected, this dipole would project onto the very largest scales of our reconstructed density and velocity fields, biasing measurements of bulk flows, power on ultra-large scales, and any cosmological signal that relies on isotropy. By shifting every redshift in our catalogues into the CMB rest frame, we ensure that observed anisotropies genuinely reflect cosmic structure rather than our local motion. This is done following the procedure outlined in \cite{2018A&A...612A..31P}. First, we define the line‐of‐sight unit vector in ICRS coordinates as $\hat{\mathbf{r}}(\alpha,\delta) = (\cos\alpha\,\cos\delta, \sin\alpha\,\cos\delta, \sin\delta)$,
where $\alpha$ and $\delta$ are the right ascension and declination (in radians) for each object in our catalogues. We then adopt the Sun’s motion relative to the CMB dipole from \cite{2007ASPC..379...24T}, expressed in ICRS, as
\begin{align}
\mathbf{v}_{\rm CMB}
& = \bigl(v_x,\;v_y,\;v_z\bigr) \\
& = \bigl(-360.5226,\;76.3969,\;-45.3542\bigr)\ \mathrm{km\,s^{-1}}.
\end{align}
Hence, the line‐of‐sight component of this velocity is given by the dot product
\begin{equation}
v^{\rm CMB}_{\rm LOS}(\alpha,\delta)
= \mathbf{v}_{\rm CMB}\,\cdot\,\hat{\mathbf{r}}(\alpha,\delta).
\end{equation}
Finally, each observed redshift $z_{\rm obs}$ is corrected to the CMB frame via 
\begin{equation}
z_{\rm CMB}
= \frac{z_{\rm obs} + v^{\rm CMB}_{\rm LOS}(\alpha,\delta)/c}
       {1 - v^{\rm CMB}_{\rm LOS}(\alpha,\delta)/c}\,,
\end{equation}
where $c = 299~792.458\;\mathrm{km\,s^{-1}}$. 

Lastly, we note that although subsequent \textit{Quaia} analyses use the $G < 20.5$ sample \citep[see][]{2023JCAP...11..043A,2024MNRAS.527.8497M,2024PhRvD.110j3025O,2025arXiv250210168A}, they do so without applying the stricter completeness cuts described above. In this work, we analyse both the \textit{Quaia Clean} and \textit{Quaia Deep Cut} samples independently.  By applying identical reconstruction pipelines to both samples, we directly assess the impact of these cuts and thereby test the robustness of our framework.  Table~\ref{tab:table_of_runs} provides a complete overview of all runs, including sample definitions and grid resolutions.

\begin{figure}
	\centering
        \includegraphics[width=\columnwidth]{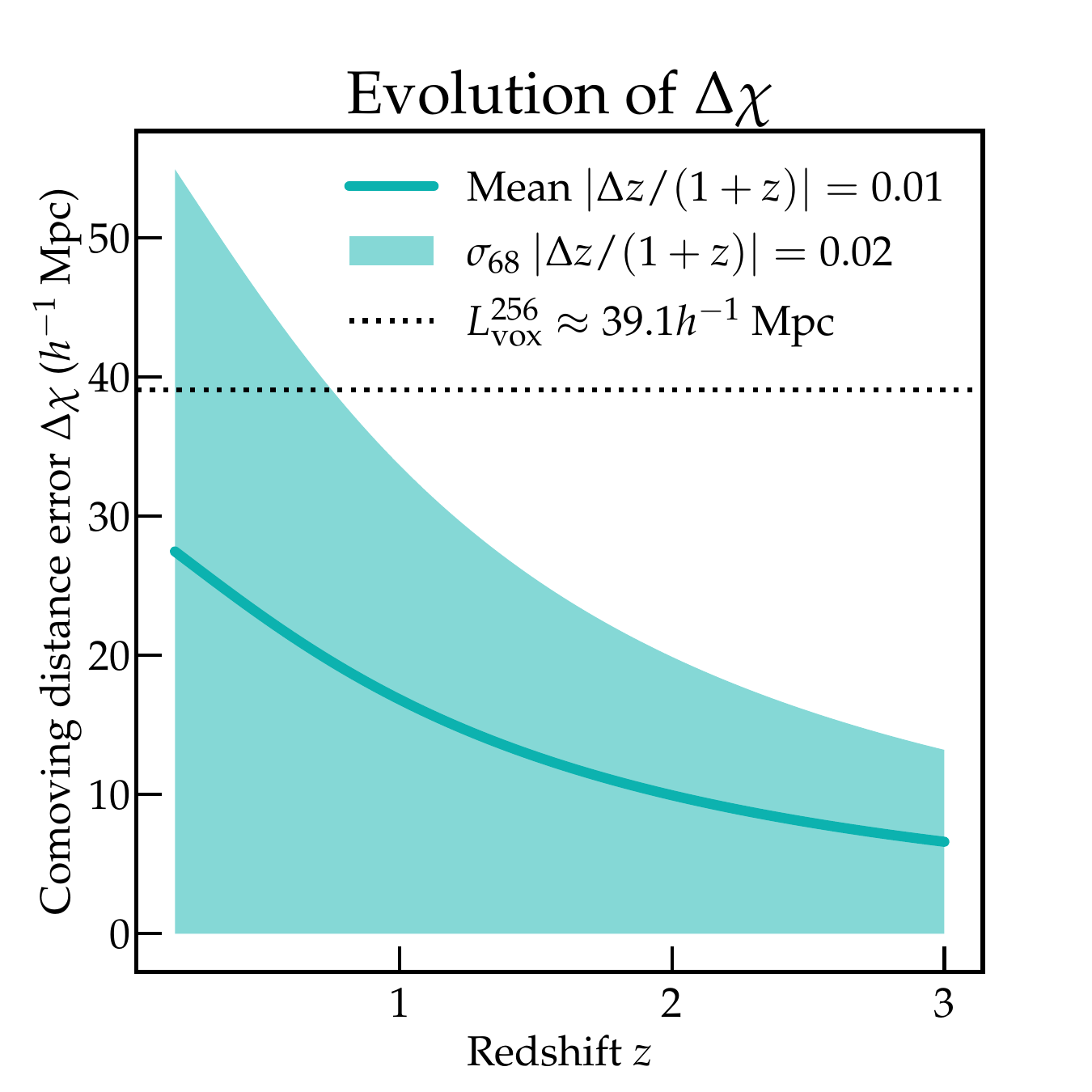}
	\caption{Evolution of the comoving distance mean error as a function of redshift compared to the smallest scale considered in this work, $L_{\rm vox}^{256}$. This shows that, on average, the error on the SPZ redshifts is well below the voxel resolution and will not affect our reconstructions.}
	\label{fig:deltachi_v_z}
\end{figure}

\label{list_of_runs}
\begin{table*}
\centering
\begin{tabular}{|l|c|c|c|c|c|}
\hline
\textbf{Index} & \textbf{Name} & \textbf{Voxel Resolution}  & \textbf{$k_{\rm max}$} &\textbf{Data Sample} & \textbf{Note} \\ \hline
1 & \textit{Quaia Deep Cut} N256    & $39.1\, \Mpch$  & $0.08$ \hMpc & $G<20.5$ & No LMC \& SMC, completness $>0.5$ \\  \hline
2 & \textit{Quaia Deep Cut} N128     & $78.1\, \Mpch$ & $0.04$ \hMpc & $G<20.5$ & No LMC \& SMC, completness $>0.5$  \\ \hline
3 & \textit{Quaia Clean} N256     & $39.1\, \Mpch$  & $0.08$ \hMpc  & $G<20.0$ &  \\ \hline
4 & \textit{Quaia Clean} N128     & $78.1\, \Mpch$ & $0.04$ \hMpc & $G<20.0$ &  \\ \hline
5 & \textit{Quaia Deep} N128      & $78.1\, \Mpch$ & $0.04$ \hMpc & $G<20.5$ & Used in Appendix \ref{Appx:SystContamination} \\ \hline

\end{tabular}
\caption{Overview of the simulation runs analysed in this work. We consider both the "\textit{Deep Cut}" and "\textit{Clean}" samples at high resolution with a box size of $L_{\rm box} ~39\, \Mpch$. Our fiducial results are based on the \textit{Quaia Deep Cut} N256 run, though we also present some results for the \textit{Quaia Clean} N256 run. Additionally, we include a lower-resolution \textit{Quaia Clean} N128 run as a resolution study for the results presented in Section \ref{sec:cmb_cross}. In this paper, we define $k_{\rm max}$ as the Nyquist frequency, $k_{\rm NF}$, of the grid.
}
\label{tab:table_of_runs}
\end{table*}

\subsection{Spectrophotometric redshift errors}\label{sect:SPZ}

The \textit{Quaia} catalogues adopt a hybrid spectrophotometric (SPZ) redshift estimation method designed to balance accuracy, precision, and a low outlier fraction. This method leverages \textit{Gaia}-derived redshifts while incorporating machine learning corrections from a \textit{k}-nearest neighbours (\textit{k}NN) approach trained on spectroscopic redshifts from SDSS-IV \citep{2020ApJS..250....8L} and unWISE \citep{2014AJ....147..108L,2019PASP..131l4504M}. The details are provided in Section~3.2 of \cite{2024ApJ...964...69S}. The key steps in the methodology are as follows:

\begin{itemize}[labelwidth=\widthof{\textbullet},leftmargin=\dimexpr\labelwidth+\labelsep\relax]
    \item if the \textit{Gaia} redshift estimate is already highly precise, meaning that the deviation satisfies  $|\Delta z / (1+z)| < 0.05$, then the original \textit{Gaia} redshift is retained: $z_{\rm \textit{Quaia}} = z_{\rm gaia}$,
    
    \item if the \textit{Gaia} estimate is likely unreliable, with a significant discrepancy, $|\Delta z / (1+z)| > 0.1$, then the \textit{k}NN redshift from SF24 is adopted: $z_{\rm \textit{Quaia}} = z_{\rm knn}$. 
\end{itemize}
Here, $|\Delta z| = |z_{\rm \textit{Quaia}} - z_{\rm spec,SDSS}|$ represents the absolute difference between the \textit{Quaia} redshift and the corresponding SDSS spectroscopic redshift. The k-Nearest Neighbours redshift, $z_{\rm kNN}$, presented in SF24, is derived by combining \textit{Gaia} photometry and unWISE mid-infrared data with \textit{Gaia} spectral features that are cross-matched to SDSS DR16 quasar spectra. For intermediate cases, a smooth transition function is applied between $z_{\rm gaia}$ and $z_{\rm kNN}$ to avoid discontinuities and ensure a continuous, well-behaved redshift distribution.

For sources with available spectroscopic redshifts from SDSS, the \textit{Quaia} $G < 20.5$ sample has a mean error of  $\langle |\Delta z / (1+z)| \rangle = 0.01$ and a 68\% scatter of 0.02, indicating that the vast majority of sources have redshift errors well below typical photometric redshift uncertainties. This performance demonstrates that the SPZ redshifts in \textit{Quaia} are significantly more reliable than typical photo-\textit{z}s, reducing catastrophic outliers ($|\Delta z / (1+z)| > 0.2$) by a factor of 3, and improving the overall redshift precision, particularly at fainter magnitudes. 

However, we are particularly interested in assessing the impact of the reported SPZ redshift errors on our field-level reconstructions. To quantify this, we propagate the redshift uncertainty, $\Delta z$, into comoving distances $\chi$ by:  
\begin{equation}
    \Delta \chi \approx \frac{\textrm{d}\chi}{\textrm{d}z} \frac{\Delta z}{(1+z)} = \frac{c\Delta z}{H(z)(1+z)}, 
\end{equation}  
where $H(z)$ is the Hubble function \citep{1996MNRAS.282..877B}.

Using the mean redshift error and dispersion reported by SF24, we compute the corresponding comoving distance uncertainties, which are shown in Figure~\ref{fig:deltachi_v_z}. For comparison, we also display the voxel size for a $N=256$ resolution grid. Our results indicate that, for the highest resolution considered in this study ($N=256$, see Table~\ref{tab:table_of_runs}), the SPZ errors remain, on average, smaller than the voxel size, $L_{\rm vox}^{256}\approx 39.1 \, h^{-1}\textrm{Mpc}$. 
Additionally, tests using mock catalogues without SPZ errors show that the observed suppression of small-scale features is primarily driven by shot noise from the low tracer density. By varying the number density in these mocks, we verified that shot noise dominates over any residual effects that could plausibly be attributed to photometric redshift uncertainties, confirming that SPZ effects are subdominant at the resolutions considered here. Consequently, the redshift uncertainty does not seem to significantly impact the reconstructions presented in Section~\ref{sec:results}.

An important mitigating factor is the high shot noise resulting from the low number density of quasars in the \textit{Quaia} catalogue. This sets a characteristic scale in the reconstruction below which density fluctuations cannot be reliably recovered. Crucially, this scale is larger than the scale at which SPZ uncertainties would impact the analysis. In other words, the noise inherent in the sample effectively washes out small-scale structure before SPZ errors become significant. We confirmed this behaviour using self-consistent mock catalogues that do not include SPZ errors: the impact of shot noise dominates the reconstruction at scales where photo-$z$ effects would otherwise appear. Combined with our chosen voxel resolution (see Figure~\ref{fig:deltachi_v_z}), this ensures that SPZ uncertainties are subdominant in our analysis. This conclusion is further supported by the validation results presented in Section~\ref{sec:cmb_cross}.

\section{Method}

In this section, we outline the methodological framework used to reconstruct the large-scale structure traced by the \textit{Quaia Deep Cut} sample. Our overarching goal is to characterise the observed structure in a physically consistent manner by inferring the primordial initial conditions from which it arose. We begin by introducing the \borg{} algorithm, a Bayesian forward-modelling approach that infers initial conditions as well as late-time matter and velocity fields by exploring the high-dimensional posterior. We then describe the forward model used to evolve primordial fluctuations into present-day observables, including the generation of initial conditions, the use of Lagrangian Perturbation Theory for gravitational evolution, and the treatment of survey selection effects. Finally, we summarise the analysis configurations employed to test the robustness of the reconstruction across data samples, resolutions, and bias models. Additional technical details -- including the survey window construction, convergence tests, the Hamiltonian Monte Carlo sampler, and the likelihood and bias model -- are provided in Section \ref{Appx:TechDets}.

\label{sec:method}
 \begin{figure}
	\centering
        \includegraphics[width=\columnwidth]{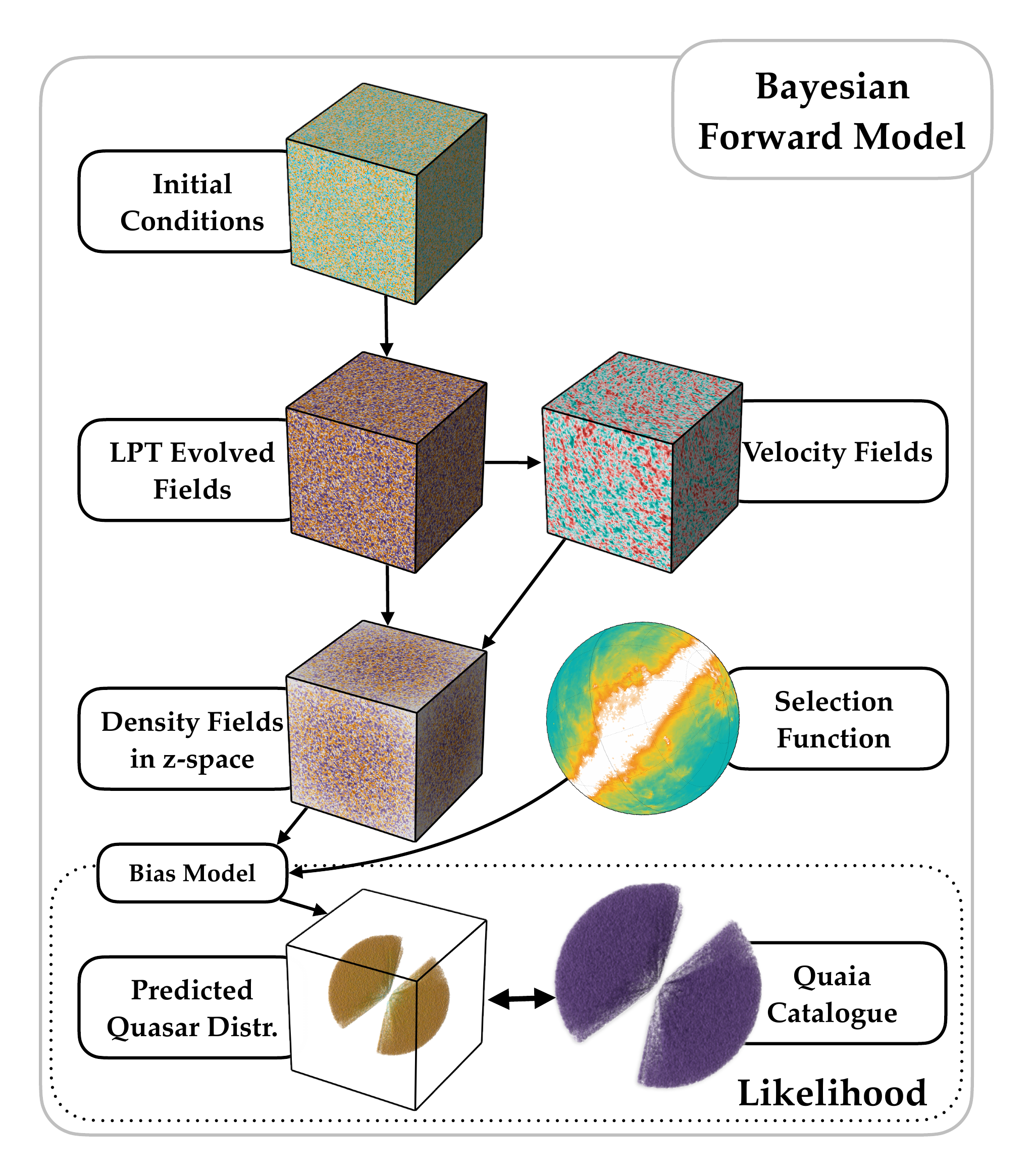}
	\caption{Flow chart of the forward model adopted in this paper, summarising the steps from initial conditions to late-time observables as described in Section \ref{ssect:forward_model}. The model begins with a Gaussian white-noise field, which is modulated by the primordial power spectrum to generate the initial gravitational potential. The resulting density field is evolved using Lagrangian Perturbation Theory (LPT) to predict matter and velocity fields at late times. Observational effects, including redshift-space distortions, light-cone effects, and survey selection functions, are then applied to yield the final predicted galaxy distribution.}
	\label{fig:borg_fc}
\end{figure}
\begin{figure*}
	\centering
    \includegraphics[width=2\columnwidth]{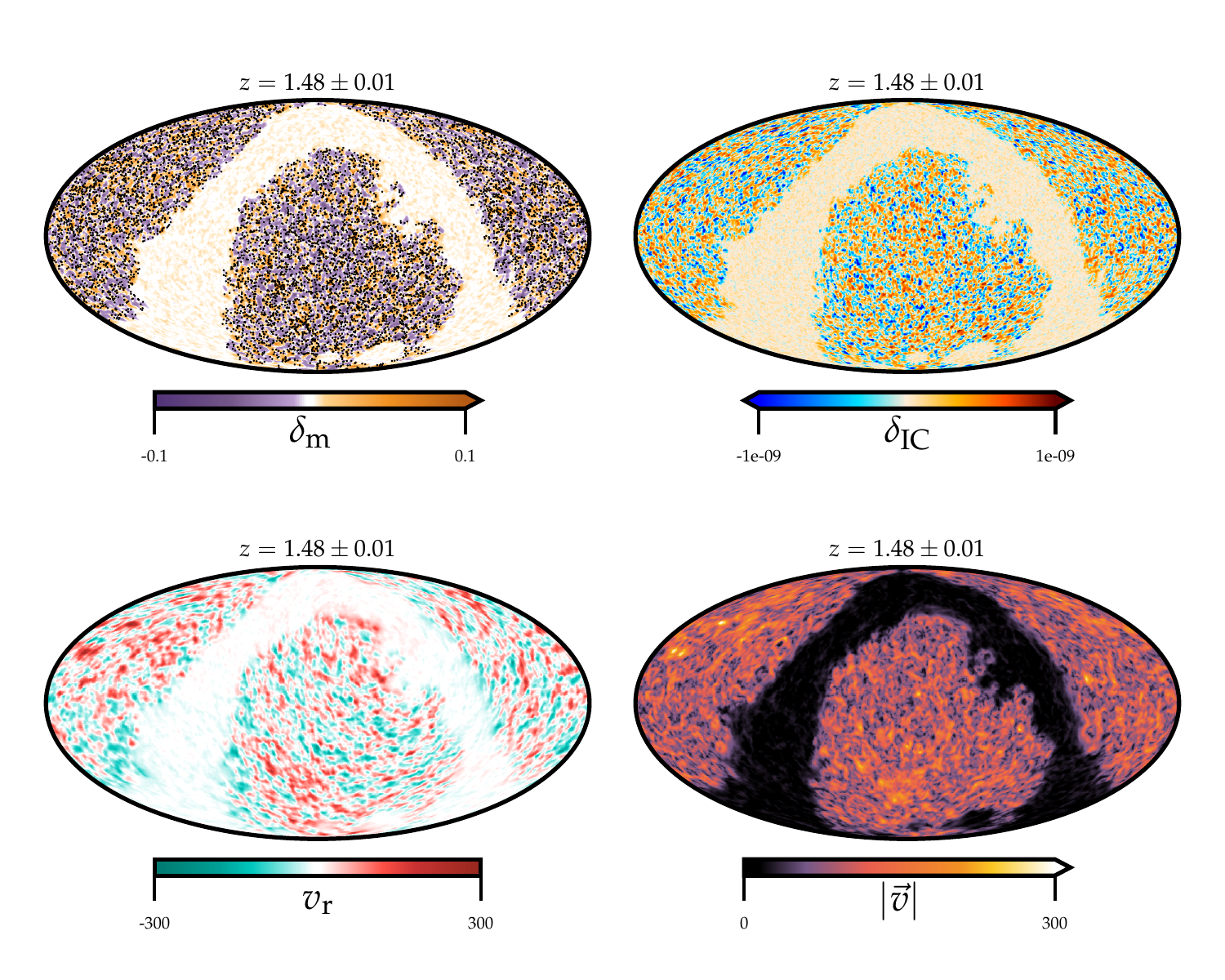}
	\caption{
    Visualisation of the inferred fields of \textit{Quaia Deep Cut} at the median redshift $z=1.48$, with a thickness of $\Delta z = 0.01$, shown in Mollweide projection. Each panel displays the mean field computed from the ensemble of MCMC samples. The top-left panel shows the inferred dark matter density field, $\delta_{\rm m}$, with overlaid quasars indicating the original input data. The top-right panel presents the inferred initial conditions, $\delta_{\mathrm{IC}}$. The bottom-left panel displays the radial velocity field, $v_{\rm r}$, while the bottom-right panel shows the absolute velocity, $|\mathbf{v}|$, computed from the full velocity vector field. The imprint of the survey footprint is evident in the inferred fields, showcasing how \borg{} is able to account for the observational systematic effects through its forward-modeling approach. These figures highlight \borg{}'s ability to infer the underlying 3D fields from quasars, providing access to the entire inferred cosmic volume along with the full statistical ensemble of MCMC realisations. This enables robust uncertainty quantification and facilitates further analysis beyond this visualisation example.}    
	\label{fig:radial_gallery}
\end{figure*}

\begin{figure*}
	\centering
    \includegraphics[width=2\columnwidth]{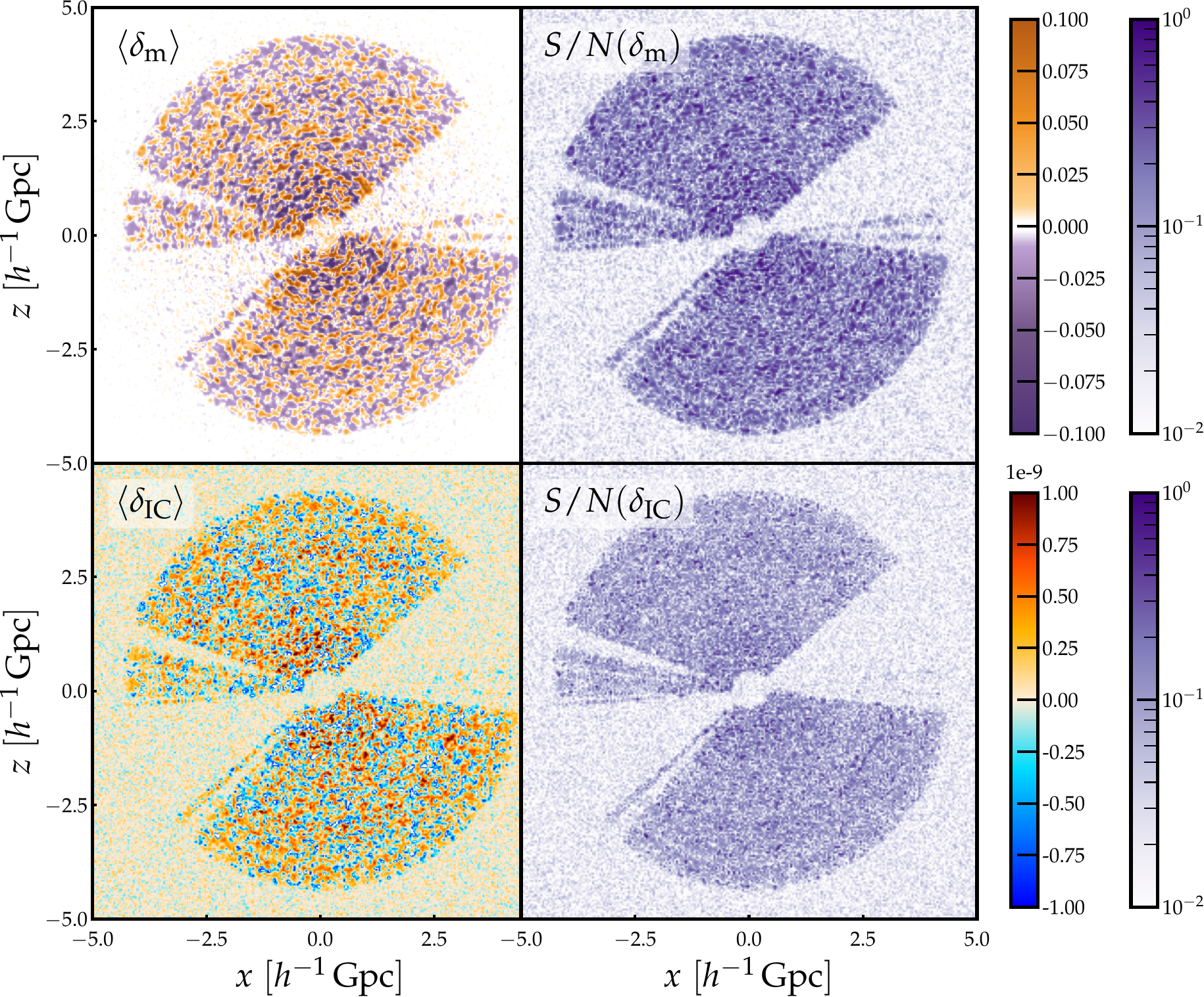}
	\caption{
        Multipanel visualisation of the inferred density fields and initial conditions for the \textit{Quaia Deep Cut} sample at a resolution of N256. Each panel shows a slice through the field at $y=0$, with a depth of $39 \, \Mpch$. The top-left panel presents the mean inferred density field, which visualizes the clumpy structure and distinctive gravitational features of the large-scale matter distribution at the present day. The top-right panel shows its corresponding Signal-to-Noise Ratio (S/N), demonstrating that the strongest observational constraints are localized in the high-density regions (clusters and filaments) where the quasar tracers are most abundant. The bottom-left panel displays the mean inferred initial conditions, which are smoother and Gaussian, reflecting the fact that the reconstruction effectively reversing the non-linear effects of gravitational evolution. The bottom-right panel illustrates their S/N, confirming that the most reliable constraints on the initial conditions are extracted from the largest, most coherent scales in the survey volume. The imprint of the survey window function is displayed through the increased uncertainty and the recovery of the cosmic mean outside the survey boundaries.
    }
	\label{fig:overlay_gallery}
\end{figure*}

\begin{figure*}
	\centering
        \includegraphics[width=2\columnwidth]{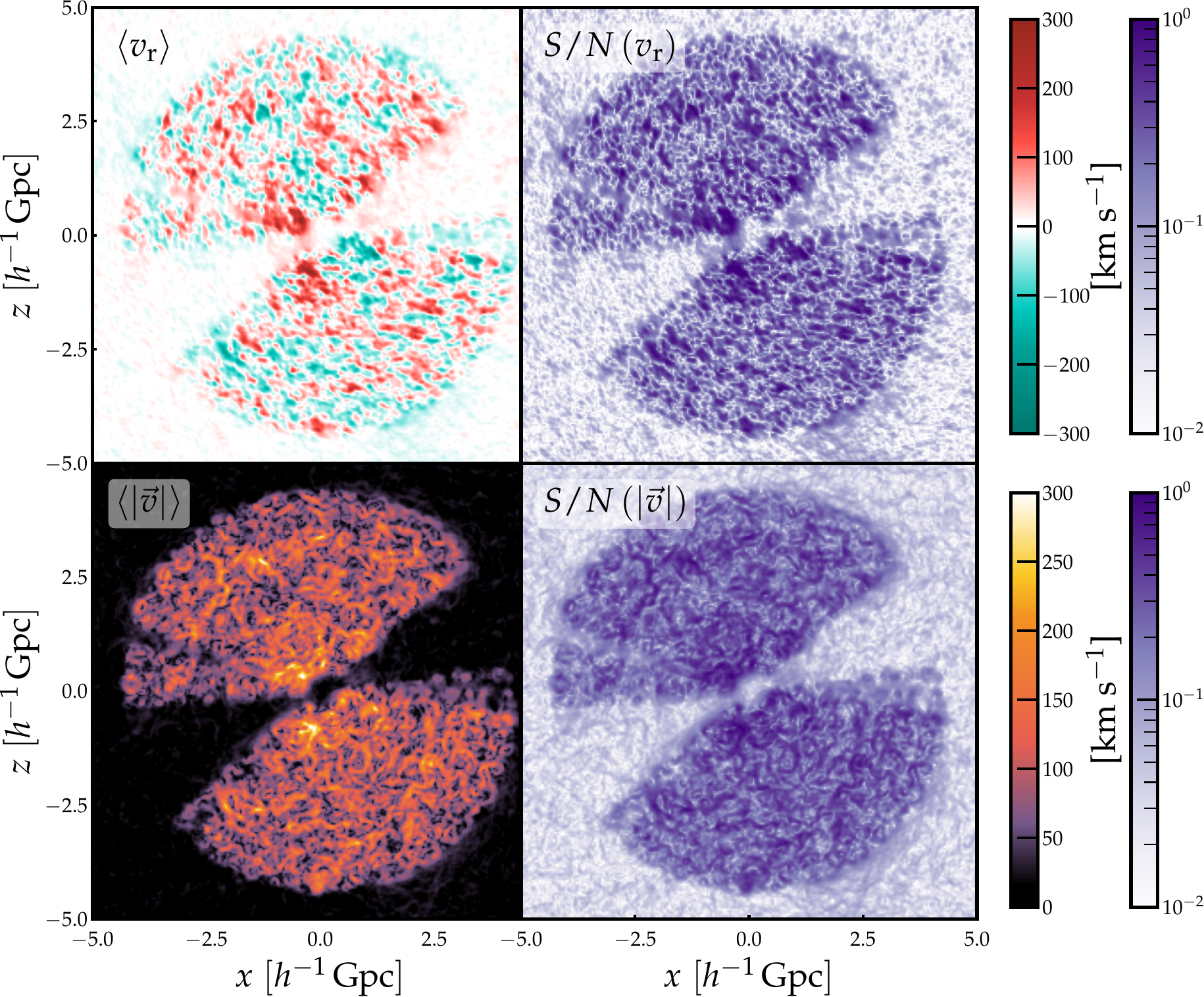}
	\caption{Similar to Figure \ref{fig:overlay_gallery}, but for the radial velocity field $v_{\rm r}$ and the absolute velocity field $|\vec{v}|$ for the \textit{Quaia Deep Cut} sample. The top-left panel presents the mean inferred radial velocity, while the top-right panel shows its corresponding S/N. The bottom-left panel displays the mean inferred absolute velocity, and the bottom-right panel illustrates its S/N.}
	\label{fig:mean_std_vel_gallery}
\end{figure*}
\subsection{\borg{} Algorithm}
The \borg{} (Bayesian Origin Reconstruction from Galaxies) algorithm is a Bayesian hierarchical forward-modeling framework for inferring the initial conditions of the Universe from observed galaxy distributions. Essentially, \borg{} uses a physical forward model to translate an arbitrary set of initial conditions to a final, predicted dataset. To explore this high-dimensional parameter space, spanned by the physics forward model, \borg{} employs a statistical sampling approach, which combines the Hamiltonian Monte Carlo algorithm and slice samplers. By utilizing the Hamiltonian Monte Carlo algorithm to efficiently explore the posterior distribution of the initial conditions, \borg{} is able to generate plausible samples given observed data, together with corresponding realisations of the present-day dark matter field. We depict the forward model adopted in this work in Figure~\ref{fig:borg_fc}. For comprehensive technical details of the \borg{} algorithm and its applications, we refer the interested reader to the literature, e.g. \cite{2010MNRAS.407...29J,2012MNRAS.425.1042J,2013MNRAS.432..894J,2016MNRAS.455.3169L,2015JCAP...01..036J,2016MNRAS.455.3169L,2019A&A...625A..64J,2019arXiv190906396L,2025MNRAS.540..716M}.

For each Markov Chain Monte Carlo (MCMC) sample, we obtain physically plausible realisations of (i) the initial conditions, (ii) the present-day dark matter field, (iii) velocity fields, and (iv) the predicted quasar distribution. These inferred fields allow us to compute a range of summary statistics, including statistical moments, two- and three-point correlation functions, and posterior predictive observables. By analysing the full MCMC chain, we can also derive ensemble averages and associated uncertainties, yielding a fully probabilistic characterisation of the large-scale structure. As an illustration, we present a slice of the inferred fields in Mollweide projection at $z = 1.48 \pm 0.01$ in Figure~\ref{fig:radial_gallery}. We assume a fixed cosmology \citep{2020A&A...641A...6P}: $\Omega_{\rm m}=0.3111$, $\Omega_{\rm b}=0.04897$, $\Omega_{\rm k}=0.0$, $\Omega_{\Lambda}=0.6889$, $n_{\rm s}=0.9665$, $A_{\rm s}=2.105\times10^{-9}$, $h=0.6766$, and $w=-1$.

\subsection{Forward Model}
\label{ssect:forward_model}

\begin{figure*}
	\centering
        \includegraphics[width=2\columnwidth]{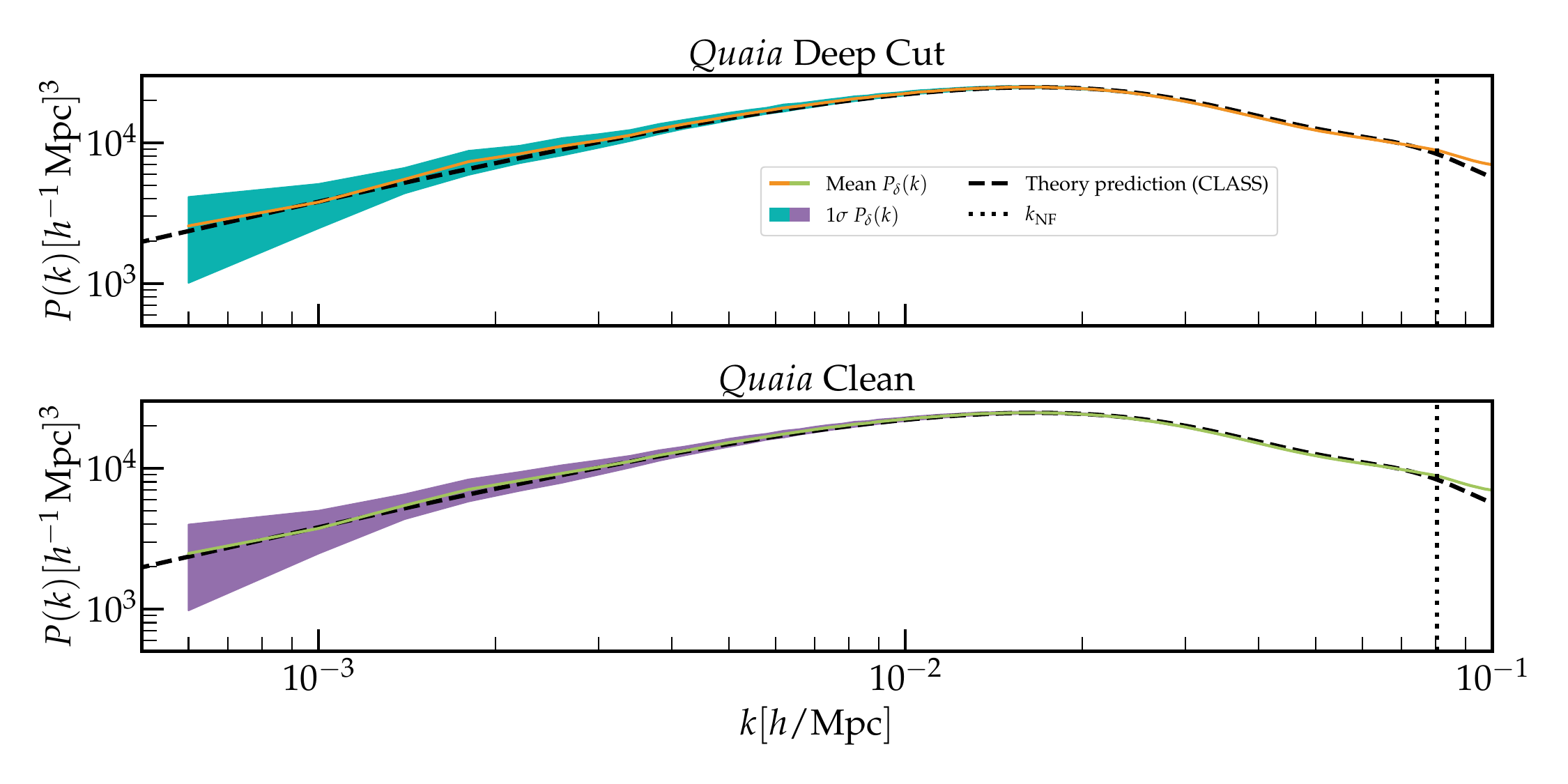}
	\caption{
        The inferred matter power spectrum $P_\delta(k)$, with uncertainty estimates, compared to theoretical predictions. 
        The solid lines show the mean $P_\delta(k)$ of the full simulation boxes, with the shaded regions representing the $1\sigma$ uncertainties. 
        The dashed black line corresponds to the linear theory prediction calculated using the \texttt{CLASS} Boltzmann code. 
        The vertical dashed line indicates the Nyquist frequency $k_{\rm NF}$, marking the smallest scale resolved by the simulation. 
        This plot emphasizes the agreement between the inferred power spectrum and theoretical expectations across a range of frequencies $k$, highlighting the validity of the inference.
    }
    \label{fig:full_pk}
\end{figure*}

The \borg{} algorithm employs a physical forward model of cosmic structure formation that iteratively reconstructs the evolution of density fluctuations from the initial conditions to the present-day observations. Below, we provide a short summary of the design choices for the forward model:

\begin{itemize}
    \item The forward model begins by generating a three-dimensional Gaussian white-noise field. This field is then convolved with a specified primordial power spectrum, to generate the primordial gravitational potential $\phi_{\rm g}$. The primordial power spectrum is defined by the assumed cosmological parameters, and sets the initial conditions for the subsequent structure formation.
    \item Next, the initial density field is convolved with a linear transfer function, computed using the Cosmic Linear Anisotropy Solving System (\texttt{CLASS}, \citealt{2011arXiv1104.2932L,2011JCAP...07..034B}). This transfer function describes the evolution of density perturbations under gravitational influence during the linear regime, and sets the initial conditions for the gravity solver. 
   \item We model gravitational collapse and subsequent structure formation using Lagrangian Perturbation Theory (LPT, \citealt{1970A&A.....5...84Z}). Given the large scales and relatively coarse resolution ($39.1\;h^{-1}\,\mathrm{Mpc}$ per voxel) of our N256 reconstruction, first-order LPT provides a computationally efficient description of the density evolution from the initial conditions into the mildly non-linear regime \citep{2013MNRAS.432..894J}. This step yields predictions for the matter density and velocity fields at late times.
    \item Finally, we also include observational effects that impact galaxy surveys, e.g. redshift-space distortions and light-cone effects \citep{2019arXiv190906396L}. We model the survey window function and selection effects through the survey response operator, $\mathcal{R}_{p}$. This operator encodes the survey footprint, completeness, and radial selection function (Figure \ref{fig:20_5_mask} and \ref{fig:20_5_rsf}), and it is applied at the voxel level.
\end{itemize} 
A schematic overview of the forward model is presented in Figure \ref{fig:borg_fc}. We refer the interested reader to the Appendix \ref{Appx:TechDets} for more details on the algorithm.

We note that relativistic corrections are not part of the forward model, as their impact on the large scales considered here is expected to be at the percent level \citep{2014CQGra..31w4004R}. With our forward model and at the resolution considered, these contributions remain well below the dominant sources of uncertainty. As a result, we expect that the analysis remains robust without the inclusion of these relativistic terms.

\subsection{Analysis Setups}
We carried out multiple reconstruction runs to test the robustness of our results against variations in sample selection, resolution, and bias modelling. First, we analysed both the \textit{Quaia Clean} and \textit{Quaia Deep Cut} samples. For each sample, we explored two bias prescriptions -- a Poisson power-law model (see Section \ref{sect:lh_and_bias}) and a Gaussian linear bias model \citep{2023MNRAS.520.5746A} -- and two analysis resolutions ($\Delta L = 78.1 \Mpch$ and $\Delta L = 39.1 \Mpch$). Although both bias models produced qualitatively similar density and velocity fields, we find the Poisson power-law bias to be more flexible and stable across scales; accordingly, all main results presented here use this model. By comparing reconstructions at different resolutions and with alternative bias schemes, we verify that our inferred fields are insensitive to these choices.

Our fiducial reconstruction is the \textit{Quaia Deep Cut} at $N_{\rm side}=256$ using the Poisson power-law bias model.  Table~\ref{tab:table_of_runs} lists all other runs shown in this work, detailing for each data sample (\textit{Clean}, or \textit{Deep Cut}), and grid resolution. This overview highlights our baseline configuration alongside the alternative setups used to validate our results.

\begin{figure*}
	\centering
        \includegraphics[width=\textwidth]{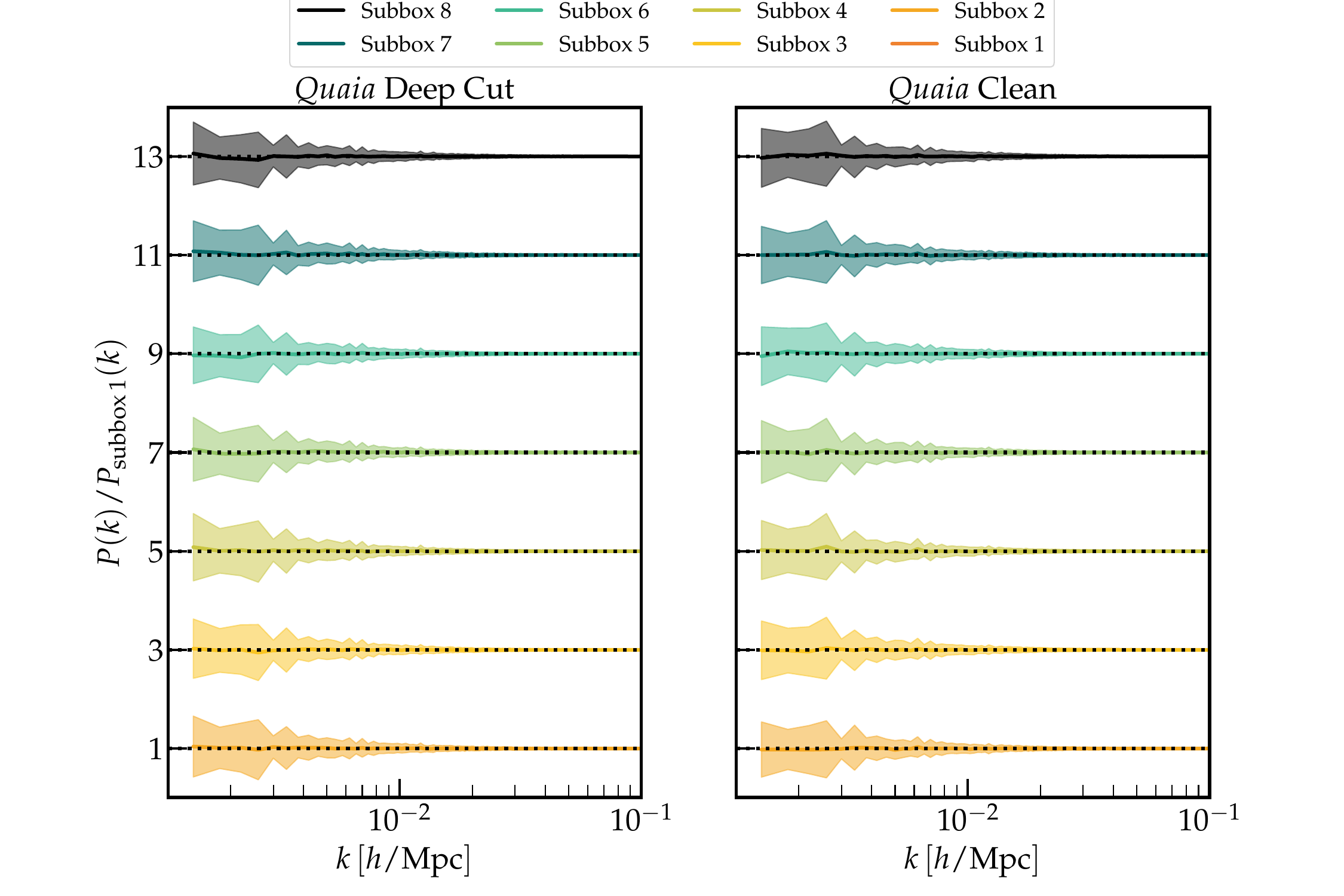}
	\caption{Visualisation of the relative deviations of the power spectra $P_\delta(k)$ measured in individual subboxes, normalised to the power spectrum of subbox~1. Each solid line shows the mean ratio $P_\delta^{(i)}(k) / P_\delta^{(1)}(k)$ for a given subbox, with the corresponding $1\sigma$ uncertainty indicated by the shaded region. For clarity, the curves are vertically offset, with the lowest displayed subbox centred at unity. This plot highlights the consistency and variance of subbox power spectra over a range of wavenumbers $k$.
    }
    \label{fig:div_plot}
\end{figure*}

\section{Inferred Fields}
\label{sec:results}

\begin{figure*}
	\centering
        \includegraphics[width=2\columnwidth]{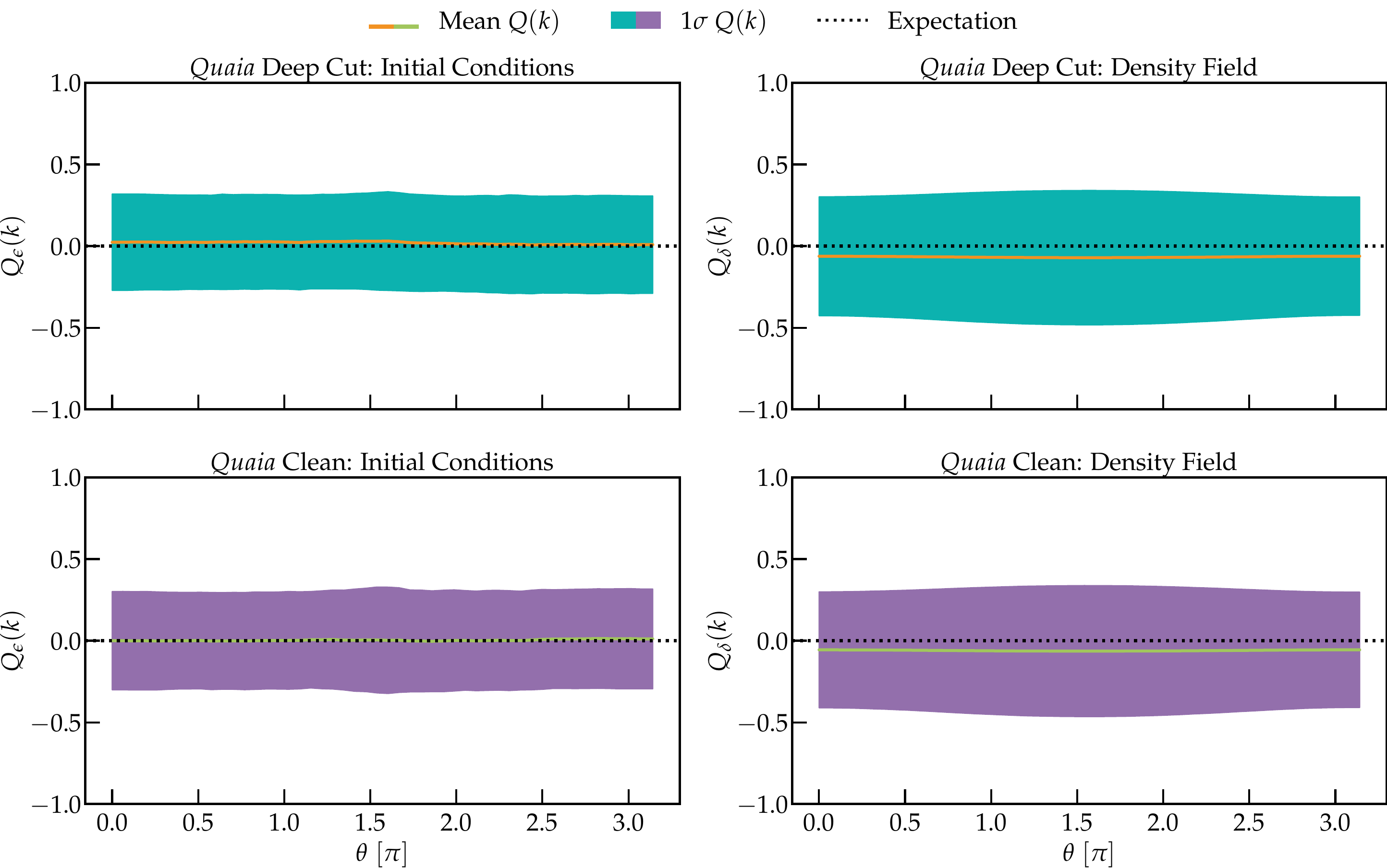}
	\caption{
    Visualisation of the reduced bispectra of the initial conditions $Q_{\epsilon}(k)$ (top) and present-day dark matter field $Q_{\delta}(k)$ (bottom) for an ensemble of inferred samples from the analysis of \textit{Quaia Clean} (left) and \textit{Quaia Deep Cut} (right). The horizontal dashed lines mark the expected reduced bispectrum $B(k) = 0$. The reduced bispectra is computed between $k_{\rm min}=0.000660 \hMpc$ and $k_{\rm max}=0.076 \hMpc$. The inferred $Q(k)$ suggests that \borg{} reconstructs the initial conditions and density field within the expected regime.}

	\label{fig:bk_ensemble}
\end{figure*}

In this section, we present and showcase the inferred fields of the \textit{Quaia} data using the \borg{} framework. The main outputs of our inference include the initial conditions, the present-day dark matter density field, and the large-scale velocity field. These data products provide a comprehensive three-dimensional view of the underlying matter distribution and its dynamical properties. In particular, the velocity field is inferred as a vector field, capturing the bulk motions of matter, and reflects the physical large-scale gravitational potential. We visualize the density field, initial conditions, and radial and absolute velocity fields in Figure \ref{fig:radial_gallery}, shown in Mollweide projection at $z \approx 1.48$, the median redshift of the \textit{Quaia} survey.

Our inference is based on a fully probabilistic approach, where we explore the posterior distribution of these fields through the MCMC samples generated. This allows us to recover not only the mean of the posterior distribution, but also quantify uncertainties in the reconstruction. While we primarily visualise the ensemble-averaged mean and standard deviation of the inferred fields, our methodology grants access to the full posterior distribution, as permitted by the data, enabling higher-order statistical analyses beyond the voxel-wise point estimates presented here.

\subsection{Matter Density Field and Initial Conditions}
\label{results_inferred_dm_ic}

The reconstruction of the dark matter density fields and primordial initial conditions is performed within a cubic volume of $V = (10 \, \Gpch)^3$, with a resolution down to $\Delta L= 39 \, \Mpch$, which means that it primarily captures large-scale, linear structures. As a result, the inferred field does not resolve the non-linear features of the cosmic web, such as filaments, sheets, and clusters, which emerge from resolving non-linear gravitational structure formation. Instead, the recovered density field provides a smoothed representation of the underlying matter distribution, which is well-suited for studying linear-scale cosmological signals.

In the upper panels of Figure \ref{fig:overlay_gallery}, we present a slice through the inferred dark matter density field. The left panel shows the ensemble-averaged mean density contrast, while the right panel displays the corresponding Signal-to-Noise Ratio (S/N), highlighting regions of varying uncertainty. We define S/N as: $\mathrm{S/N} = \frac{\langle X \rangle}{\sigma(X)}$, for the field value $X$. Overlaid on the mean field are the quasars from the \textit{Quaia} survey, which trace the overdense regions, demonstrating qualitative consistency between the inferred density field and the observed large-scale distribution of tracers.  

In the bottom panels of Figure \ref{fig:overlay_gallery}, we display the inferred initial conditions at the corresponding slice. The left panel shows the ensemble-averaged mean initial conditions, while the right panel presents the corresponding S/N. As expected, there is a clear visual correlation between the large-scale features in the initial conditions and the inferred density field, with overdense regions in the present-day matter distribution tracing peaks in the initial fluctuations.

Our inference framework naturally incorporates selection effects through its forward-modelling approach, with survey systematic effects included as part of this treatment \citep[as described in][]{2024ApJ...964...69S}, as reflected in the recovered density and initial condition fields. In regions lacking direct observational constraints, both fields approach the cosmic mean while exhibiting increased uncertainty, indicating that the inference remains unbiased and does not introduce artificial structures due to incomplete coverage. This behaviour highlights the robustness of the Bayesian framework, which ensures self-consistent reconstructions that link present-day density fluctuations to their primordial origins.

\begin{figure}
	\centering
        \includegraphics[width=1.0\columnwidth]{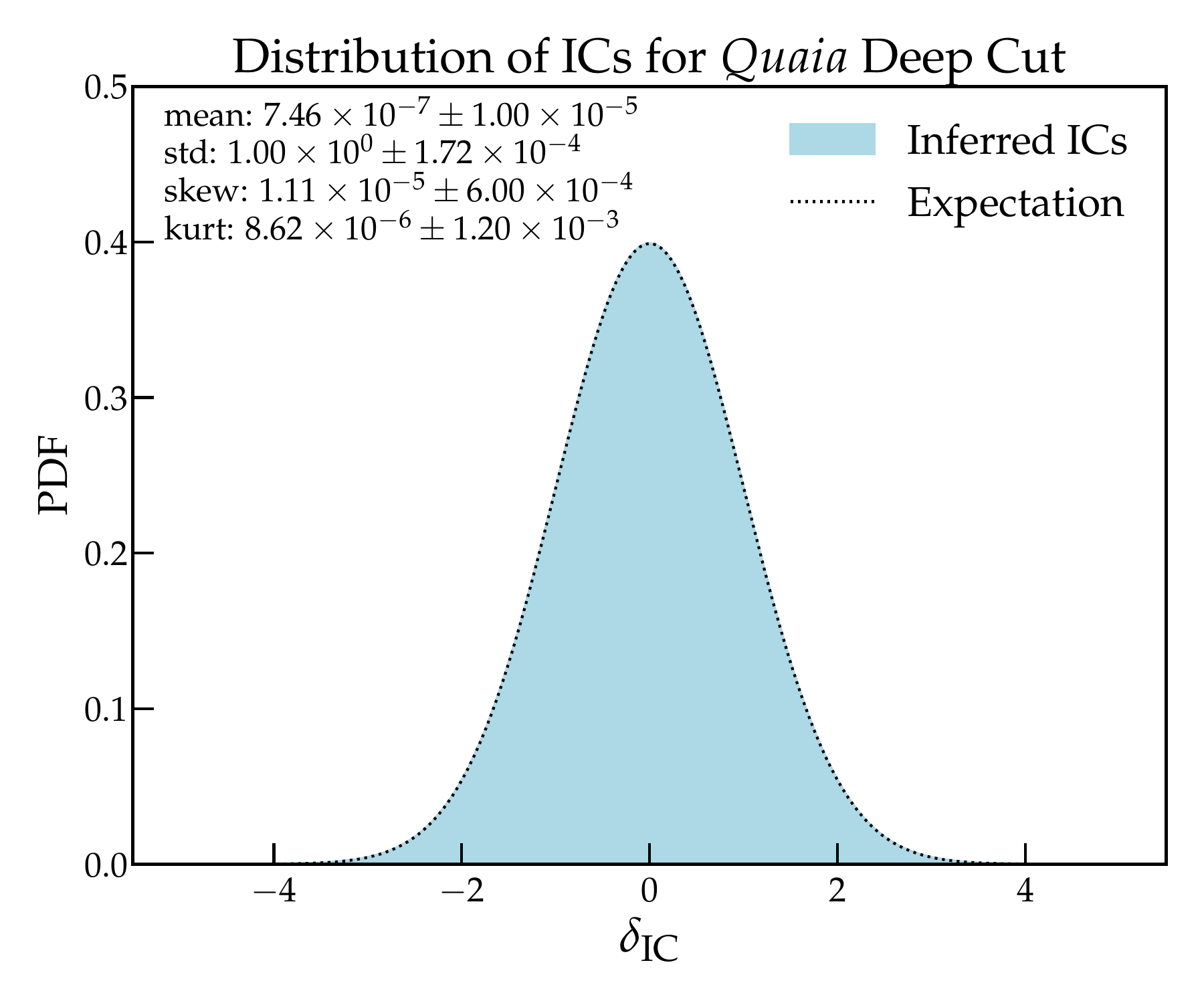}
	\caption{Distribution of the reconstructed initial conditions from the \textit{Quaia Deep Cut} catalog (blue), shown together with the theoretical Gaussian expectation derived from the prior covariance (black, dashed). The probability density function (PDF) of the inferred field matches the Gaussian prediction, as reflected by the mean, variance, skewness, and kurtosis values listed in the inset. This agreement demonstrates that the inference combined with the forward model produces unbiased reconstructions, i.e. no statistically significant deviations from Gaussianity are present beyond expectations. The result validates the sufficiency of the statistical inference framework in capturing the data at the level of one-point statistics.}
    \label{fig:IC_hist}
\end{figure}

In summary, our inference of the dark matter density field and initial conditions demonstrates consistency with the observed quasar distribution while accounting for survey systematics. More broadly, the recovered fields capture structures across a wide range of scales, offering a physically motivated representation of the matter distribution traced by quasars. The inferred initial conditions provide a valuable data product for follow-up studies, for example constrained simulations \citep{2022MNRAS.512.5823M, 2022MNRAS.509.1432S, 2024A&A...691A.348W}. In the Section \ref{sec:validation}, we present the recovered 2-pt and 3-pt statistics of these fields, and assess their agreement with theoretical expectations.

\subsection{Velocity Field Reconstruction}

The primary goal of this section is to analyse the three-dimensional velocity field derived from our inferred initial conditions. By computing the line-of-sight velocity field, $v_{\rm r}$, and the absolute velocity field, $|\vec{v}|$, we can visualise the large-scale dynamics of the observed volume. This reconstruction allows us to probe the statistical properties of the large-scale velocity field. 

The velocity field is obtained by forward-modelling the inferred initial conditions and extracting particle positions and velocities from the LPT output. The velocity components relative to an observer at the centre are decomposed into radial and tangential directions along the line of sight. To construct gridded fields, both the particle momenta and masses are assigned to a three-dimensional uniform cubic lattice ($N = 256$, $L = 10{,}000\,\Mpch$) using a CIC scheme, and the velocity in each cell is computed as the ratio of the CIC-assigned momentum to the CIC-assigned mass. This yields mass-weighted velocity fields suitable for visualisation.

Figure \ref{fig:mean_std_vel_gallery}, presents a visualisation of the inferred velocity fields. Each panel shows a slice through the field at $y=0$, with a depth of $39 \, \Mpch$. The figure shows the mean radial velocity field, $\langle v_r \rangle$, the corresponding S/N, $S/N(v_r)$, the mean absolute velocity, $\langle |\mathbf{v}| \rangle$, and the $S/N(\mathbf{v})$.

The inferred velocity fields reveal coherent large-scale structures, reflecting the gravitational evolution of matter within the observed volume. The radial velocity field exhibits dipolar features, with inflows and outflows tracing the underlying density distribution. The uncertainty in $v_r$ follows the survey window function, with increased uncertainty in regions with limited quasar coverage. The absolute velocity field highlights the kinematic properties of the inferred structures, with higher velocities concentrated in denser regions. The radial velocity field traces the inflow and outflow of the underlying density field, while the absolute velocity field shows kinematic properties of the inferred structures, e.g., higher velocities concentrated in denser regions. Uncertainties follow the survey window function, increasing near the survey boundaries, highlighting that the analysis incorporates the full statistical ensemble of the algorithm. These results demonstrate the ability of the inference framework to recover the large-scale velocity field while accounting for observational limitations, providing a statistically plausible inference of the large-scale bulk flows.

In summary, our analysis yields the full 3D posterior distribution of the radial velocity field, enabling both detailed mapping and statistical characterisation beyond the mean reconstruction. 

\section{Validation}
\label{sec:validation}
\begin{figure*}
	\centering
        \includegraphics[width=2.0\columnwidth]{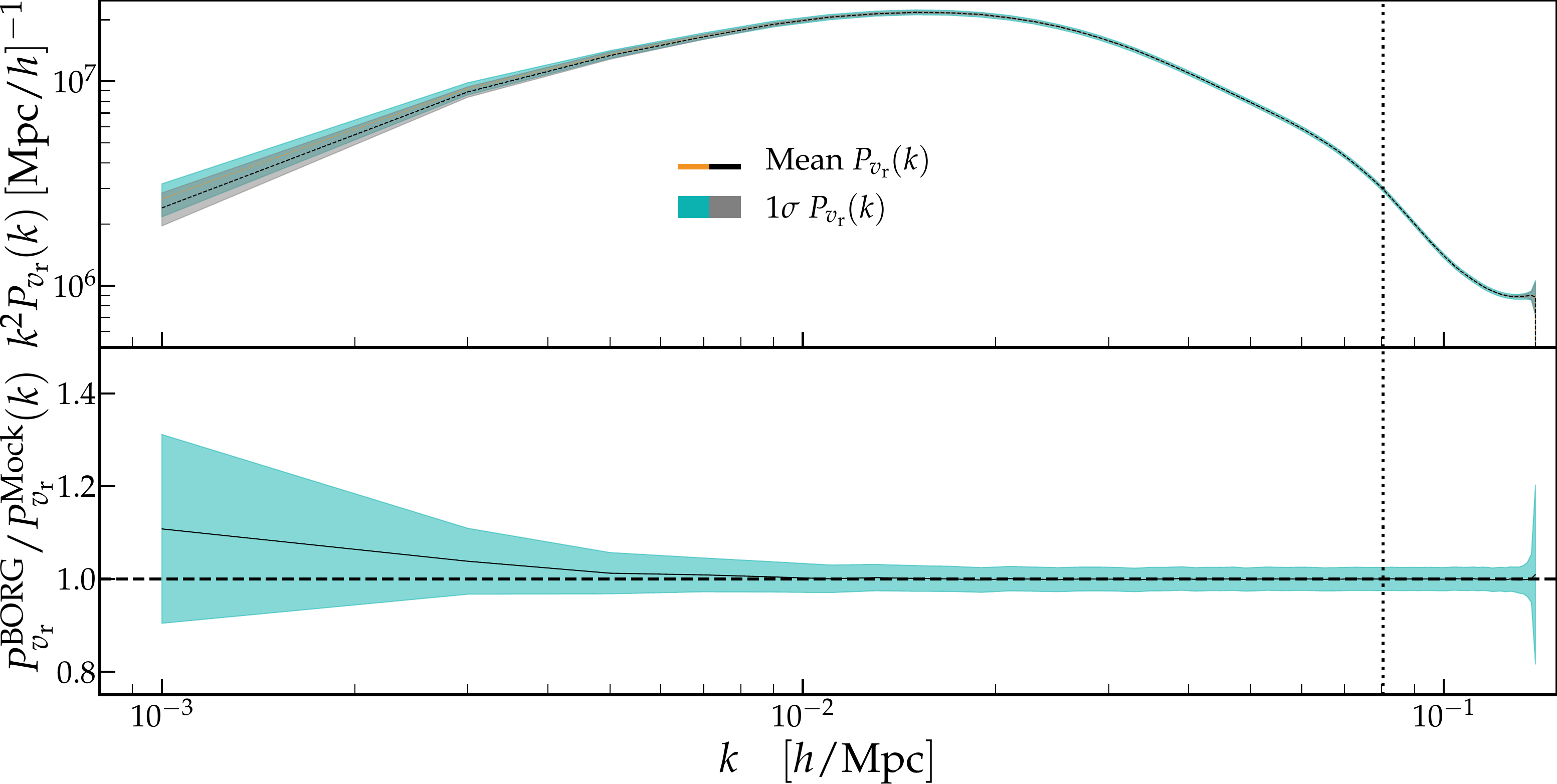}
	\caption{Power spectrum of the velocity field, $P(k)$, as a function of wavenumber $k$. The dotted line and grey shaded region represent the mean and standard deviation, respectively, from the data. The dashed blue line and light blue shaded region correspond to the mean and standard deviation from mock catalogs. The close agreement between the data and mocks indicates that the inferred velocity fields accurately captures the expected statistical properties.}

    \label{fig:vr_pk}
\end{figure*}

To ensure the robustness of our inferred data, we perform a series of validation tests on the reconstructed fields and derived data products. These include consistency checks with theoretical models of structure formation, focusing on the recovered matter power spectrum, bispectrum, and velocity fields. We also carry out posterior predictive tests, such as cross-correlations with independent observables like CMB lensing from \citet{2020A&A...641A...8P}, to verify that the inferred structures trace real features of the Universe. These validation steps are critical not only to assess the internal consistency of our approach, but also to demonstrate that the \borg{} framework reliably reconstructs the large-scale structure from biased and noisy tracer data. Ultimately, this enables us to move from raw catalogues of point sources to physically meaningful reconstructions that can be used to address fundamental questions in cosmology.

\subsection{The inferred power spectra and bispectra}

To further quantify the statistical properties of the inferred fields, we compute the power spectrum $P(k)$ and bispectrum $B(k)$ of the matter density field and the initial conditions. By computing $P(k)$ and $B(k)$ for each sample in our posterior ensemble and taking the ensemble average, we obtain estimates of the two- and three-point statistics while preserving full uncertainty information, providing complementary tests of our reconstruction. The power spectrum and bispectrum allows us to test for non-Gaussian features, and to validate our inference against theoretical predictions. In the following, we present and interpret these results. Moreover, we compare the computed power spectra of the radial velocity field with the expectation in the form of random mock data, in order to assess the consistency of our model. 

Figure~\ref{fig:full_pk} presents the ensemble-averaged power spectrum $P(k)$ of the inferred matter density field at $z=0$, computed without light-cone effects or redshift-space distortions. The figure displays results for both the \textit{Quaia Clean} (bottom panel) and \textit{Quaia Deep} (top panel) samples, illustrating the consistency of our inference across different survey selections. The range of scales covered in our analysis extends from large, linear scales to quasi-nonlinear regimes, demonstrating the ability of our method to recover structure across a broad range of scales. Furthermore, our inferred power spectra show close agreement with theoretical predictions from \textsc{CLASS}, reinforcing the reliability of the reconstruction.

To test the spatial consistency of our reconstruction, we compute power spectra in eight subregions of the data cube. Specifically, we extract initial condition fields from the MCMC samples and divide them into subvolumes of half the box size. For each subcube, we compute the corresponding power spectrum and normalise it by the power spectrum of the first subcube, as shown in Figure~\ref{fig:div_plot}. This approach allows us to assess the robustness of the inference across different regions of the reconstructed volume, ensuring that the inferred initial conditions remain statistically consistent throughout the survey footprint. Any significant deviations between subvolumes and the reference power spectrum would indicate potential contaminations, or spatial model mismatch, or other variations affecting the quality of the inference.

We note that the subcubes are not treated as independent periodic volumes in a physical sense. Rather, the power spectra are used purely as estimators to probe relative spatial consistency within the reconstructed volume. While the estimator implicitly assumes periodicity at the cube level, this assumption is applied uniformly across all subregions. Any bias introduced by this approximation therefore affects all subboxes in a similar way and largely cancels out when comparing their power spectra

We visualise the non-Gaussian features in our reconstruction through computations of the bispectra. The present-day matter fields $\delta_{\rm m}$ and the initial conditions $\epsilon$ are extracted from the MCMC samples. The reduced bispectra of these fields are then computed using \texttt{Pylians3} \citep{Pylians}, with $k_{\min}$ and $k_{\max}$ set at $3.30\times 10^{-4}\hMpc$ and $7.64\times 10^{-2}\hMpc$, respectively. The bispectrum is computed for multiple triangular configurations defined by a range of angles $ \theta $, over 50 bins. Figure~\ref{fig:bk_ensemble} compares the ensemble-averaged bispectrum from our inference with theoretical predictions. The close agreement shows that, given the LPT forward model, the method recovers the expected bispectrum without introducing spurious non-linear contributions, and is therefore consistent with zero.

In summary, the inferred two- and three-point statistics offer a way to assess whether the reconstructed fields are plausibly accurate. The power spectrum analysis confirms that the inferred density field is statistically consistent with theoretical expectations across a wide range of scales, while the bispectrum demonstrates that our inference does not introduce artificial non-Gaussianities. Moreover, the spatial consistency of the sub-box power spectrum indicates that our approach across is consistent over the inferred volume. These results highlight the ability of our method to recover large-scale statistical properties while accounting for observational systematics. In the next section, we extend our analysis to the inferred velocity fields, further probing the dynamical information contained within the reconstruction.

\subsection{1-pt statistics of the inferred initial conditions}

Figure \ref{fig:IC_hist} presents the normalised histogram of the inferred initial conditions, constructed from samples of the posterior distribution. The distribution is centred around zero, with a standard deviation of unity, as expected for a Gaussian field. Additionally, the skewness and kurtosis of the inferred initial conditions are consistent with zero, indicating that no significant non-Gaussian features are introduced by the inference process. These results demonstrate that the forward model used in the inference is sufficient to describe the data generation process and that our data model does not introduce systematic biases, reinforcing the validity of our approach in reconstructing the initial conditions from the observed quasar distribution.

\subsection{Validation of velocity fields}

To compare with theoretical predictions, we generate an empirical reference by constructing and analysing mock velocity fields using Lagrangian perturbation theory. For each random mock realisation, we compute the radial velocity field from the particle velocities. Repetition of this process over an ensemble of $1000$ mock fields provides a statistical estimate of the expected velocity fields, providing a reference for comparison with our inferred results.

To test the inferred velocity fields, we compare the power spectra of the inferred fields with the power spectra of the mock samples in Figure \ref{fig:vr_pk}. The dotted line and the shaded grey region represent the mean and standard deviation of the inferred velocity field. Additionally, the comparison with mock catalogues (dashed blue line and light blue shaded region) demonstrates the consistency of the reconstructed velocity field with theoretical expectations. The agreement across scales indicates that the inference procedure accurately captures the statistical properties of the velocity field, reinforcing the reliability of the method.

Validation against theoretical predictions confirms the expected two-point correlation, demonstrating the reliability of our approach; additional tests and results are presented in the appendix (Appendix \ref{Appx:velocity_dipole}).

\section{Cross-correlation with CMB Lensing}\label{sec:cmb_cross}


Validating \borg{} dark matter density field reconstructions presents a significant 
challenge in field-level inference, especially when aiming to ensure that the inferred 
structures accurately reflect the true distribution of matter in the Universe. 
The lensing of the cosmic microwave background (CMB) provides an invaluable dataset 
for such validation \citep{2019arXiv190906396L}, as it directly probes the integrated matter distribution along the line of sight. 
If the dark matter density field inferred by \borg{} samples truly represents 
the actual large-scale structure of the Universe within the scales probed by the data, 
then it should exhibit measurable correlations with the weak gravitational lensing 
imprints observed in the CMB by \textit{Planck} \citep{2020A&A...641A...8P}.

To test the fidelity of the \borg{} field-level reconstruction of the matter density field,
$\delta_{\rm m}(\textbf{r})$, on the largest scales accessible via the \textit{Quaia} quasar sample, 
we perform a  cross-correlation analysis. 
Specifically, we measure the correlations between the weak 
lensing convergence derived from \borg{}'s reconstructed matter density field, 
$\kappa_{\rm BORG}$, with the \textit{Planck} 2018 Lensing Convergence map, $\kappa_{\rm \textit{Planck}}$. 
This approach leverages the statistical power of cross-correlation to ascertain whether the
reconstructed structures align with the lensing-induced distortions of the CMB, offering a
robust validation of the \borg{} data products. This validation not only confirms the 
accuracy of the \borg{} reconstruction at the largest scales probed by \textit{Quaia}, but also 
strengthens our understanding of the sample, giving us confidence for future 
cosmological analyses and applications.

\subsection{\borg{} Convergence Maps}

The forward model implemented in \borg{} provides samples of the reconstructed underlying matter density field,
$\delta^{\rm f}_{\rm m} (\mathbf{r}) = \delta_{\text{m}}^{\rm final}(\chi, \hat{n})$, defined in a light-cone and convolved with a Cloud-In-Cell (CIC) kernel \citep{hockney_eastwood,2008ApJ...687..738C, 2009arXiv0901.3043J}. The CIC kernel in Fourier space is given by:
\begin{equation}
    W(\mathbf{k}) = \prod_{i=1}^3 \left\{ \frac{\sin (\pi k_i / 2 k_{\rm Ny})}{\pi k_i / 2 k_{\rm Ny}}\right\}^2,
    \label{eq:CIC}
\end{equation}
where $k_{\rm Ny}$ is the Nyquist wavenumber, and $k_i (i=1,2,3)$ are the components of the wavenumber vector $\mathbf{k}$. To accurately represent the matter density field, we perform a deconvolution of the CIC kernel in Fourier space, yielding
\begin{equation}
    \delta_{\text{m}}^{\rm f, dec}(\mathbf{k}) =
    \begin{cases}
        W^{-1}(k)\,\delta_{\text{m}}^{\rm f}(\mathbf{k}), & \text{if } |W(k)| > \epsilon, \\
    0, & \text{otherwise},
    \end{cases}
\end{equation}
where, we used $\epsilon=10^{-12}$. Although the CIC deconvolution is not exact, this step ensures that the reconstructed field does not contain correlation artifacts introduced by the CIC grid. 

Following the procedure outlined in Section 4.3 of \cite{2019arXiv190906396L}, the 3D convergence field we obtain the matter density lensing convergence, $\kappa(\hat{n})$,  by integrating the matter density field along the line of sight with the lensing kernel, up to the comoving distance to the last scattering surface, $\chi_{\rm CMB}$,
\begin{equation}
    \kappa(\hat{n}) = \frac{3 \Omega_{\rm m} H_0^2}{2 c^2} \int_0^{\chi_{\rm CMB}} \frac{\chi}{a(\chi)} \left( \frac{\chi_{\rm CMB}^{} - \chi}{\chi_{\rm CMB}^{}} \right) \delta_{\text{m}}^{\rm f, dec}(\chi, \hat{n}) \, \mathrm{d}\chi,
\end{equation}
where $a(\chi)$ is the scale factor at comoving distance $\chi$, $H_0$ is the Hubble constant in km/s/$h^{-1}$Mpc, $\Omega_{\rm m}$ is the matter density at present time and $c$ is the speed of light. Note that we are assuming a flat Universe in this work, with the angular comoving distance expressed as $f_{\rm K}(\chi) = \chi$.

The next step involves projecting the 3D convergence field onto a \texttt{HEALPix} map with a resolution of $N_{\rm side} = 2N_{\rm vox}^{1/3}$. This is achieved using a tri-linear interpolation scheme\footnote{See \url{https://paulbourke.net/miscellaneous/interpolation/} for details. Although this interpolation introduces some degree of smoothing along the line-of-sight, the voxel-to-pixel effects dominate. We have verified this effect has no measurable via a resolution study for both integration and \texttt{HEALPix} resolutions.} to perform the line-of-sight integration, ensuring accurate mapping from the Cartesian grid to the spherical map. The final product, $\kappa_{\rm BORG}$, is a \texttt{HEALPix} map of the lensing convergence, encapsulating distortions induced by the reconstructed matter density field as traced by the \textit{Quaia} quasar sample over the redshift range $0.174 < z < 3.0$ with a source at the CMB. 

\subsection{\textit{Planck} PR3 CMB Lensing Map}\label{sect:PlanckLensing}

For this analysis, we used the publicly available CMB lensing convergence map from the 2018 
release (PR3) of the \textit{Planck} satellite, as described by
\cite{2020A&A...641A...8P}\footnote{\url{https://pla.esac.esa.int/pla/}}.
Specifically, we use the minimum variance (MV) convergence map, $\kappa_{\rm CMB}$, which combines 
information from spin-0 and spin-2 CMB fields to minimise variance. As recommended by the \textit{Planck} 
Collaboration, we filter out modes with $\ell < 8$ to reduce large-scale systematic contamination 
and improve robustness. 

The \textit{Planck} lensing maps are provided in harmonic space. Hence, to ensure compatibility with the \borg{} 
convergence maps for cross-correlation, we transform the \textit{Planck} maps into real space at the target 
\texttt{HEALPix} resolution $N_{\rm side}$, matching that of the \borg{} maps. This step avoids numerical
biases that may arise from downgrading or interpolating the CMB lensing maps during the power spectra
analysis. Additionally, we apply the CMB lensing mask provided by \cite{2020A&A...641A...8P}, 
which was improved to reduce contamination from point sources with no loss of area.

\subsection{Angular Power Spectrum Measurements}

\begin{figure*}
    \centering
    \begin{subfigure}[t]{0.483\textwidth}
        \centering
        \includegraphics[width=\linewidth]{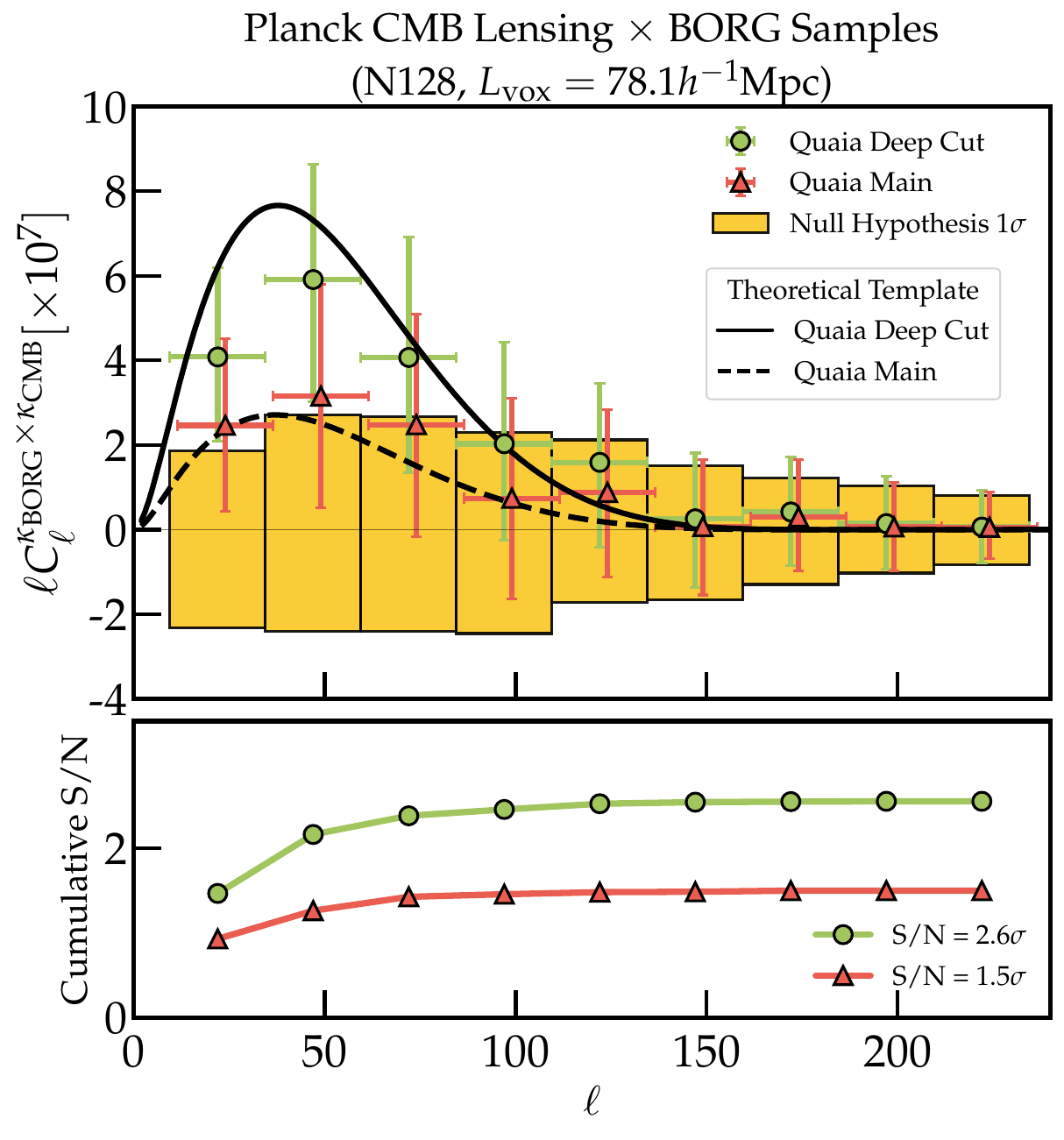}
        \caption{$N_{\rm vox}=128^3$, $N_{\rm side}=256$}
        \label{fig:cmb_x_borg_compare}
    \end{subfigure}
    \begin{subfigure}[t]{0.49\textwidth}
        \centering
        \includegraphics[width=\linewidth]{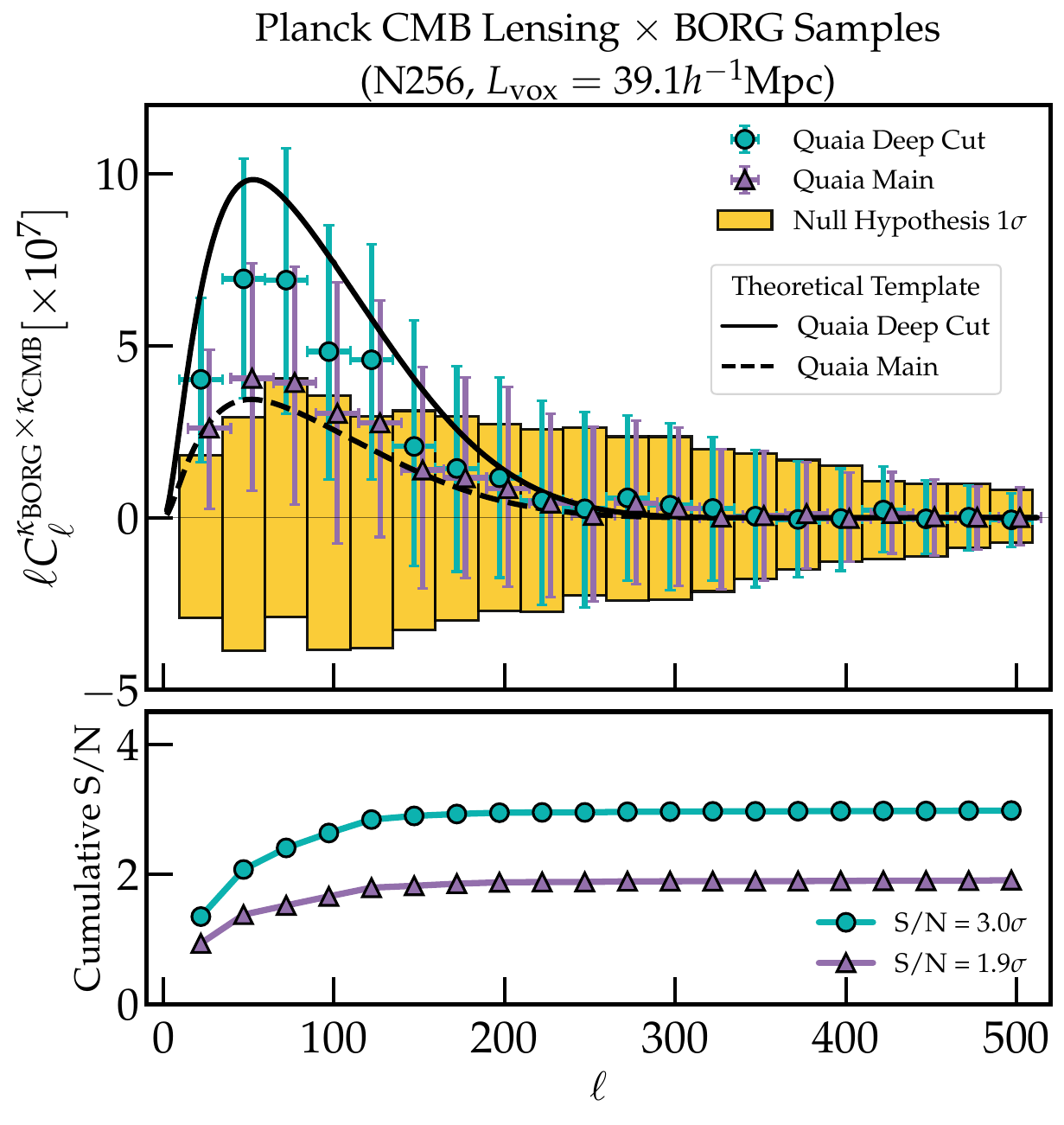}
        \caption{$N_{\rm vox}=256^3$, $N_{\rm side}=512$}
        \label{fig:cmb_x_borg_deep}
    \end{subfigure}\hfill

    \caption{\textit{Upper panels:} Cross-correlations between the \textit{Planck} CMB lensing map, $\kappa_{\rm {Planck}}$, and lensing convergence from \borg{} matter density field reconstructions,
    $\kappa_{\rm BORG}$. Circles (triangles) show the mean of the cross-correlations from 1000 \textit{Quaia Deep Cut} (\textit{Quaia Main}) \borg{} samples and their $1\sigma$ errors. The yellow boxes show the $1\sigma$ regions for the null hypothesis -- randomly oriented \borg{} reconstructions cross-correlated with $\kappa_{\rm {Planck}}$. The black lines are the theoretical templates for the signal -- \texttt{CCL} prediction convolved with the voxel-to-pixel window function (Appendix \ref{Appx:cmb_signal_modelling}) and assuming the amplitude parameter $A=1$ ( Section~\ref{sec:amplitude-fitting}). \textit{Lower panels:} Cumulative signal-to-noise ratio as a function of multipoles (Section~\ref{sec:snr} for details) for \textit{Quaia Deep Cut} (circles) and \textit{Quaia Main} (triangles). The value quoted in the legend is the final cumulative signal-to-noise ratio over the multipoles considered for each case.}

    \label{fig:cmb_xcorr_combined}
\end{figure*}

We measure the cross-angular power spectra of the \borg{} lensing convergence, $\kappa_{\rm BORG}$, and the CMB lensing convergence, $\kappa_{\rm CMB}$, using \texttt{NaMaster} \citep{2019MNRAS.484.4127A}. To account for the effects of the observational mask, we deconvolve the mixing of modes using the mixing matrix obtained from the mask. This approach avoids adding unnecessary complexity to the theoretical validation of our results, as explicitly modelling both the mask-induced mode mixing and the voxel-to-pixel window function (see Appendix \ref{Appx:cmb_signal_modelling}, Figure \ref{fig:vox2pix_window}) simultaneously would complicate the interpretation of the measurements. We leave details about the modelling of the cross-correlations signal for Appendix \ref{Appx:cmb_signal_modelling}, including the modelling of the effect of the voxel-to-pixel window function.

To construct the effective mask for our analysis, we combine the \textit{Quaia} completeness mask (Figure~\ref{fig:20_5_mask}) with the \textit{Planck} PR3 lensing mask. The combined mask is apodised using \texttt{NaMaster} with a \texttt{Smooth} apodisation scheme\footnote{For more information, see \url{https://namaster.readthedocs.io/}} and a smoothing scale of $0.15$ degrees. Apodisation reduces edge effects, ensuring smoother transitions at the boundaries of the mask. This step is crucial for enabling a numerically stable deconvolution of the mask effects and minimising contamination in the measured angular power spectra \citep{2002ApJ...567....2H}.

In our validation analysis, we compute the cross-angular power spectra between $\kappa_{\rm BORG}$ and the \textit{Planck} CMB Lensing map, $\hat{C}_{\ell}^{\kappa_{\rm B}\kappa_{\rm C}}$, for the main configurations summarised in Table \ref{tab:table_of_runs}. For the N256 (N128 respectively) resolution, the angular power spectra are computed over the multipole range $\ell_{\rm min} = 10$ to $\ell_{\rm max} = N_{\rm side} = 512$ ($256$ respectively) with a bandwidth of $\Delta\ell = 25$. 

Figure~\ref{fig:cmb_xcorr_combined} presents the mean and 68\% Confidence Interval (C.I.) for the cross-angular power spectra, $\hat{C}_{\ell}^{\kappa_{\rm B}\kappa_{\rm C}}$, between the \textit{Planck} CMB lensing map, described in Section~\ref{sect:PlanckLensing}, and 1000 samples for $\kappa_{\rm BORG}$ derived from \textit{Quaia} samples as defined in Table~\ref{tab:table_of_runs}. The figures also shows the show the theoretical signal estimated from \texttt{CCL} \citep{2019ApJS..242....2C}, as described in Appendix \ref{Appx:cmb_signal_modelling}. As expected, the \textit{voxel-to-pixel} window function (see Appendix \ref{Appx:cmb_signal_modelling}, Figure \ref{fig:vox2pix_window}) significantly suppresses the cross-power on smaller angular scales (i.e., higher multipoles) in the lower-resolution maps, shifting the effective loss of power to larger scales.

\subsubsection{Amplitude fitting and detection significance}\label{sec:amplitude-fitting}
\begin{table}
    \centering
    \caption{
    Best-fit amplitude parameters, detection significances, and phase alignment cumulative signal-to-noise ratio (S/N) for the CMB lensing cross-correlations between the \citet{2020A&A...641A...8P} convergence map and the \borg{} convergence reconstructions of \textit{Quaia} samples at different resolutions.
    }
    \label{tab:detection_significance}
    \begin{tabular}{lccc}
        \hline
        Sample (Resolution) & Amplitude, $A$ & \shortstack{Detection\\Significance}  & \shortstack{Phase Align.\\ $S/N(\ell_{\rm max})$} \\
        \hline
        \textit{Quaia Deep Cut} (N128) & $0.81 \pm 0.24$ & $ 3.34\,\sigma$ & $2.6\,\sigma$ \\
        \textit{Quaia Deep Cut} (N256) & $0.70 \pm 0.17$ & $ 3.96\,\sigma$ & $3.0\,\sigma$ \\
        \textit{Quaia Main} (N128) & $1.29 \pm 0.64$ & $ 2.03\,\sigma$ & $1.5\,\sigma$ \\
        \textit{Quaia Main} (N256) & $1.23 \pm 0.50$ & $ 2.44\,\sigma$ & $1.9\,\sigma$ \\
        \hline
    \end{tabular}
\end{table}

Following the methodology adopted in previous CMB lensing cross-correlation analyses \citep{2015PhRvD..91f2001H, 2015PhRvD..92f3517L, 2020ApJ...904..182M, 2022MNRAS.515.1993S, 2021A&A...649A.146R}, we quantify the detection significance of the CMB lensing signal by fitting a single amplitude parameter, $A$, which rescales a fixed theoretical template for the cross-angular power spectrum, $\mathbf{t} \equiv C_{\ell}^{\kappa_{\rm B}\kappa_{\rm C}}$. Under the assumption of a Gaussian likelihood for the measured binned cross-correlation data vector, this procedure allows for an analytic determination of both the maximum-likelihood amplitude, $\hat{A}$, and its associated uncertainty,$\sigma_{\hat{A}}$, from which the detection significance is directly inferred.

The theoretical template is computed using \texttt{CCL} \citep{2019ApJS..242....2C}, adopting the same \cite{2020A&A...641A...6P} cosmological parameters assumed in the \borg{} forward-modelling inference\footnote{$\Omega_{\rm m} = 0.3111,\, \Omega_{\rm b} = 0.04897,\, \Omega_{\rm k} = 0.0,\, \Omega_{\Lambda} = 0.6889,\, n_{\rm s} = 0.9665,\, A_{\rm s} = 2.105 \times 10^{-9},\, h = 0.6766,\, w = -1$.}, ensuring consistency between the signal model and the reconstructed large-scale structure.
While this approach accounts for the radial selection function of the \textit{Quaia} sample, it does not fully capture the effects of the \borg{} forward-modelling approach (e.g., bias model), and therefore the resulting template should be regarded as a theoretically motivated best-case shape for the cross-correlation, with any residual mismatches absorbed by the fitted amplitude parameter, $A$. Further details on the construction of the theoretical cross-correlation template are provided in Appendix~\ref{Appx:cmb_signal_modelling} and shown in Figure~\ref{fig:cmb_xcorr_combined} as black lines for the different \textit{Quaia} samples and \borg{} reconstruction resolutions.

Building on the Gaussian-likelihood assumption introduced above, we model the mean binned cross-angular power spectrum, $\mathbf{d} \equiv \langle \hat{C}_{\ell}^{\kappa_{\rm B}\kappa_{\rm C}} \rangle$, as a scaled version of the fixed theoretical template, $ \mathbf{t}' = A\,\mathbf{t} = A C_{\ell}^{\kappa_{\rm B}\kappa_{\rm C}}$ with $A$ as a single free amplitude parameter. The goodness-of-fit for a given value of $A$ is quantified through the $\chi^2$ statistic
\begin{equation}
\chi^2(A) =
\left(\mathbf{d} - A\,\mathbf{t}\right)^{\mathsf T}
\boldsymbol{\Sigma}^{-1}
\left(\mathbf{d} - A\,\mathbf{t}\right),
\label{eq:chi2_samples}
\end{equation}
where $\boldsymbol{\Sigma}$ is the covariance matrix of the binned cross-angular power spectrum measurements, $\hat{C}_{\ell}^{\kappa_{\rm B}\kappa_{\rm C}}$.

The null hypothesis of no cross-correlation corresponds to $A=0$, for which
\begin{equation}
\chi^2_{\rm null} \equiv \chi^2(A=0) =
\mathbf{d}^{\mathsf T}\,
\boldsymbol{\Sigma}^{-1}\,
\mathbf{d}.
\end{equation}
The best-fitting model is obtained by minimising $\chi^2(A)$ with respect to the amplitude, yielding the maximum-likelihood value $\hat A$.

The detection significance is defined through the likelihood-ratio test statistic $\Delta\chi^2 \equiv \chi^2_{\rm null} - \chi^2(\hat A)$, which, for a single fitted parameter, follows a $\chi^2$ distribution with one degree of freedom under the null hypothesis.
This allows $\sqrt{\Delta\chi^2}$ to be interpreted as a Gaussian-equivalent $N\sigma$ detection significance, fully equivalent to the ratio $\hat A/\sigma_{\hat A}$. The resulting best-fit amplitudes, uncertainties, and detection significances are
summarised in Table~\ref{tab:detection_significance}.

Despite being limited by the voxel-to-pixel window (Figure~\ref{fig:vox2pix_window}), our analysis constructs a physical model of the large-scale Universe, providing, for the first time, a joint reconstruction of the initial conditions, three-dimensional density, and velocity fields traced by quasars. The measured detection significances in Table~\ref{tab:detection_significance} confirm the robustness of these field-level inference reconstructions on the largest scales probed by the \textit{Quaia} Catalogue. In particular, increasing the reconstruction resolution leads to a systematic improvement in detection significance, with the \textit{Quaia Deep Cut} Sample at N256 resolution yielding the highest detection significance, ${\sim}4\sigma$. However, spectrophotometric redshift uncertainties primarily limit further increases in resolution, which restricts our ability to model the full three-dimensional dynamics of structure formation. In combination with the low number density of the high-volume quasar sample, these uncertainties set a practical limit on the reconstruction resolution. As a result, $L_{\rm vox} = 39.1\,h^{-1}$Mpc is the highest resolution achievable without being dominated by photo-$z$ errors (Figure~\ref{fig:deltachi_v_z}) or shot noise. We also note that the \textit{Quaia Main} sample remains limited to $\lesssim 2.5\,\sigma$ at both resolutions, reflecting insufficient signal-to-noise for a standalone analysis.

These results suggest that the \textit{Quaia Deep Cut} sample provides a robust large-scale tracer when used in combination with other deep spectroscopic datasets, enabling an increase in effective survey volume with controlled systematics and making such joint samples well suited for studies of ultra-large-scale physics, including primordial non-Gaussianity \citep[e.g.][]{2023MNRAS.520.5746A,2024arXiv241211945A}.

\subsubsection{Phase alignment signal-to-noise and null tests}\label{sec:snr}
Complementary to the amplitude-based detection significances presented in Section~\ref{sec:amplitude-fitting}, we perform a model-independent assessment of the cross-correlations signal using the full binned data vector, $\hat{C}_{\ell}^{\kappa_{\rm B}\kappa_{\rm C}}$. This approach directly probes the statistical significance of the observed phase alignment between the \borg{} convergence reconstructions and the \textit{Planck} CMB lensing map, without assuming that the signal strictly follows a fixed
theoretical template. In other words, in this Section we are assessing how different is the distribution of sample cross-correlations from that of a random alignment.

Again, we denote by $\mathbf{d} \equiv \langle \hat{C}_{\ell}^{\kappa_{\rm B}\kappa_{\rm C}} \rangle$ the mean binned cross-angular power spectrum measured from the \borg{} posterior ensemble. Null realizations, $\langle \mathbf{d}_{\rm null} \rangle$, are constructed by rotating the coordinates of the \borg{} convergence maps prior to cross-correlating them with $\kappa^{\rm CMB}$ from \textit{Planck}, thereby preserving their cosmic variance while explicitly destroying any phase alignment with the CMB lensing field. The excess signal is then defined as
\begin{equation}
    \boldsymbol{\Delta} \equiv \mathbf{d} - \langle \mathbf{d}_{\rm null} \rangle ,
\end{equation}
which reduces to $\boldsymbol{\Delta} = \mathbf{d}$ for a null ensemble with zero mean.

The covariance matrices of the posterior and null samples, $\boldsymbol{\Sigma}$ and $\boldsymbol{\Sigma}_{\rm null}$ respectively, are estimated directly from the cross-correlations. Assuming statistical independence, we define the total covariance
\begin{equation}
    \boldsymbol{\Sigma}_{\rm tot} =
    \boldsymbol{\Sigma} +
    \boldsymbol{\Sigma}_{\rm null}.
\end{equation}
Using this covariance, we define a cross-correlation signal-to-noise ratio as
\begin{equation}
    \left( S/N \right)_{\rm tot}
    =
    \sqrt{
    \boldsymbol{\Delta}^{\mathsf T}
    \boldsymbol{\Sigma}_{\rm tot}^{-1}
    \boldsymbol{\Delta}
    },
\end{equation}
which corresponds to the Mahalanobis distance of the measured cross-correlation signal from the null hypothesis. All covariance inversions are performed using the Moore--Penrose pseudo-inverse to ensure numerical stability in the presence of finite-sample noise.

Finally, we investigate the scale dependence of the detection by computing the
cumulative signal-to-noise as a function of maximum multipole $\ell_{\rm max}$.
For each $\ell_{\rm max}$, we evaluate
\begin{equation}
    \left( S/N \right)(\le \ell_{\rm max})
    =
    \sqrt{
    \boldsymbol{\Delta}_{\le \ell_{\rm max}}^{\mathsf T}
    \boldsymbol{\Sigma}_{{\rm tot},\le \ell_{\rm max}}^{-1}
    \boldsymbol{\Delta}_{\le \ell_{\rm max}}
    },
    \label{eq:snr_cum}
\end{equation}
where both the signal vector and covariance matrix are truncated to bins with
$\ell \le \ell_{\rm max}$. This cumulative statistic, shown in the lower panels of
Figure~\ref{fig:cmb_xcorr_combined}, highlights the angular scales that dominate the
total detection significance and provides an additional validation of the
large-scale origin of the measured signal.

We emphasise that the amplitude-based detection significance presented in Section~\ref{sec:amplitude-fitting} and the phase-alignment signal-to-noise defined above probe complementary aspects of the same measurement.
The amplitude fit constitutes an optimal likelihood-ratio test under the assumption that the measured cross-correlation follows a fixed theoretical template, compressing the data into a single parameter, $A$.
By contrast, the phase-alignment signal-to-noise defined in Eq.~\eqref{eq:snr_cum} operates directly in data space and quantifies the statistical distance between the posterior of observed cross-correlation vectors and the distribution of null realisations.

Specifically, the quantity $(S/N)_{\rm tot}$ corresponds to the Mahalanobis distance between the mean posterior cross-correlation and the null hypothesis, measured in the metric defined by the total covariance. As such, it tests whether the observed phase alignment between the \borg{} reconstructions and the Planck CMB lensing field is consistent with a random realisation of the null ensemble, without assuming any particular signal shape.
For the \textit{Quaia Deep Cut} sample at N256 resolution, we find $(S/N)_{\rm tot} \simeq 3$, implying that the observed phase alignment is inconsistent with random alignments at the $3\sigma$ level.

\section{Summary \& Conclusions}
\label{sec:conclusions}

In this paper, we have applied the \borg{} Bayesian inference framework to the \textit{Quaia} dataset. Leveraging the wide sky coverage and depth of \textit{Quaia}, we reconstructed the three-dimensional matter distribution, primordial initial conditions, and large-scale velocity fields over a comoving volume of $(10\,h^{-1}\,\mathrm{Gpc})^3$ at a resolution of $39.1\,h^{-1}\,\mathrm{Mpc}$. This constitutes both the first application of field-level inference to quasars and the largest cosmological reconstruction by volume to date. Our analysis used two complementary quasar samples, \textit{Quaia Clean} and \textit{Quaia Deep Cut}, accounting for angular and radial selection effects, redshift uncertainties, redshift-space distortions, quasar biasing, and light-cone effects within a fully forward-modelled, hierarchical Bayesian framework.

The primary results of our inference are posterior maps of the present-day dark matter density field, the corresponding initial conditions, and large-scale velocity fields, constructed using a physical model of large-scale structure formation. The inferred density field shows strong correlations with the observed quasar distribution and extends coherently beyond the survey boundaries through gravitational coupling. The recovered initial conditions are consistent with Gaussian statistics, while the velocity field traces large-scale coherent flows aligned with overdense and underdense regions. Statistical validation demonstrates that the reconstructed fields reproduce the expected power spectrum and bispectrum of large-scale structure without introducing spurious non-Gaussianities, and remain consistent across subvolumes. Crucially, we detect a highly significant cross-correlation between the \borg{} density field and \textit{Planck} CMB lensing, with a maximum significance of $4.0\sigma$ (via template-fitting), providing an external and independent validation of the reconstruction on the largest scales probed. Meanwhile, a phase-alignment null test shows that our reconstructions is inconsistent with random alignments with the \textit{Planck} PR3 CMB lensing map at a $3\sigma$ level for \textit{Quaia Deep Cut}.

The resolution of our reconstruction is primarily limited by shot noise and spectrophotometric redshift uncertainties, which prevent reliable inference on small scales. These limitations set the effective scale of the voxel-to-pixel window (Figure~\ref{fig:vox2pix_window}), which suppresses small-scale power in the projected field. Consequently, we cannot extend the reconstruction to higher resolution without being dominated by noise or redshift uncertainties. Despite these constraints, our results demonstrate that field-level inference can robustly extract three-dimensional information from sparse and noisy tracers, fully propagating observational uncertainties and survey systematics. Our work establishes quasars as a powerful probe for field-level cosmology and shows that probabilistic reconstructions of the matter distribution are feasible over unprecedented cosmic volumes. This enables precision studies of large-scale structure, gravitational dynamics, and cross-correlations with other cosmological probes, highlighting the potential of field-level inference as a key analysis framework for forthcoming surveys such as \textit{Euclid} and beyond. Field-level inference with \borg{} thus transforms sparse quasar observations into a statistically robust, three-dimensional map of the Universe’s large-scale structure, bridging observations and theory with unprecedented fidelity.

The reconstructed three-dimensional quasar density fields open up a range of cosmological applications beyond traditional clustering analyses. In particular, the time-evolving gravitational potential inferred at the field level provides a natural framework for modelling late-time effects such as the Integrated Sachs--Wolfe (ISW) signal. By propagating the reconstructed matter distribution across cosmic time, its imprint on the CMB can be predicted in a physically consistent manner, enabling direct cross-checks with current and future CMB observations. This study provides a validated framework and data products that lay the groundwork for further cosmological analyses, including Alcock--Paczynski tests \citep{2019A&A...621A..69R} and investigations of large-scale quasar structures \citep{2013MNRAS.429.2910C,2022MNRAS.516.1557L,2024JCAP...07..055L}.

At the same time, the large reconstructed volume and robust recovery of long-wavelength modes make this framework well suited for probing primordial physics. Parameterised extensions of the initial conditions would allow joint field-level inference of local-type primordial non-Gaussianity \citep{2023MNRAS.520.5746A,2024arXiv241211945A}, yielding constraints on \( f_{\mathrm{NL}} \) that are competitive with existing quasar-based analyses. More generally, introducing a flexible model for the primordial power spectrum within the likelihood would enable tests of the standard power-law assumption and searches for large-scale features. These possibilities underscore the broader potential of field-level inference with quasar surveys to deliver novel and complementary constraints on both late- and early-Universe physics.

\section*{Acknowledgements}
The authors gratefully acknowledge the Quaia team for making their data publicly available, and in particular Kate Storey-Fisher, David Alonso, and Giulio Fabbian for their insightful discussions and insights regarding the catalogue.
 The authors also thank Ludvig Doeser for stimulating discussions, Natalia Porqueres and Eleni Tsaprazi for discusssion on inference of photometric data with \borg{}, and Julia Stadler and Harry Desmond for valuable feedback on the manuscript. 
 
 AA acknowledges with gratitude the financial support provided by the Wenner-Gren Foundations under Grant No. WGF2025-0043. AL acknowledges support from the Swedish National Space Agency (Rymdstyrelsen) under Career Grant Project Dnr 2024-00171 and from the research project grant `Understanding the Dynamic Universe' funded by the Knut and Alice Wallenberg Foundation under Dnr KAW 2018.0067. JJ and GL acknowledge support from the Simons Foundation through the Simons Collaboration on Learning the Universe. FL and GL acknowledge financial support from the Agence Nationale de la Recherche (ANR) through grant INFOCW, under reference ANR-23-CE46-0006-01. 
This work is done within the Aquila Consortium\footnote{\url{https://www.aquila-consortium.org/}}.
This research utilised the HPC facility supported by the Technical Division of the Department of Physics, Stockholm University, with a special thanks to Mikica Kocic for technical support.

The computation and data processing in this study were enabled by resources provided by the National Academic Infrastructure for Supercomputing in Sweden (NAISS), partially funded by the Swedish Research Council through grant agreement no. 2022-06725 via Project Dnr. NAISS 2024/3-18.

We acknowledge the use of the following packages: \texttt{NumPy} \citep[]{harris2020array}, \texttt{SciPy} \citep[]{2020SciPy-NMeth}, \texttt{AstroPy} \citep[]{astropy:2013,astropy:2018,astropy:2022}, \texttt{Matplotlib} \citep[]{Hunter:2007}, \texttt{Pylians3} \citep[]{Pylians}, \texttt{COLOSSUS} \citep[]{2018ApJS..239...35D}, \texttt{NaMaster} \citep[]{2019MNRAS.484.4127A}, \texttt{CCL} \citep[]{2019ApJS..242....2C}, \texttt{camb}  \citep[]{Lewis:1999bs, Howlett:2012mh}, and \texttt{HEALPix} \citep[]{2005ApJ...622..759G}.
\section*{Data Availability}

The data underlying this article will be shared on the basis of a reasonable request to the corresponding authors.
 



\bibliographystyle{mnras}
\bibliography{literature} 




\appendix

\section{Additional Results}
\label{Appx:AddResults}
\begin{figure}
    	\centering
        \includegraphics[width=1.0\columnwidth]{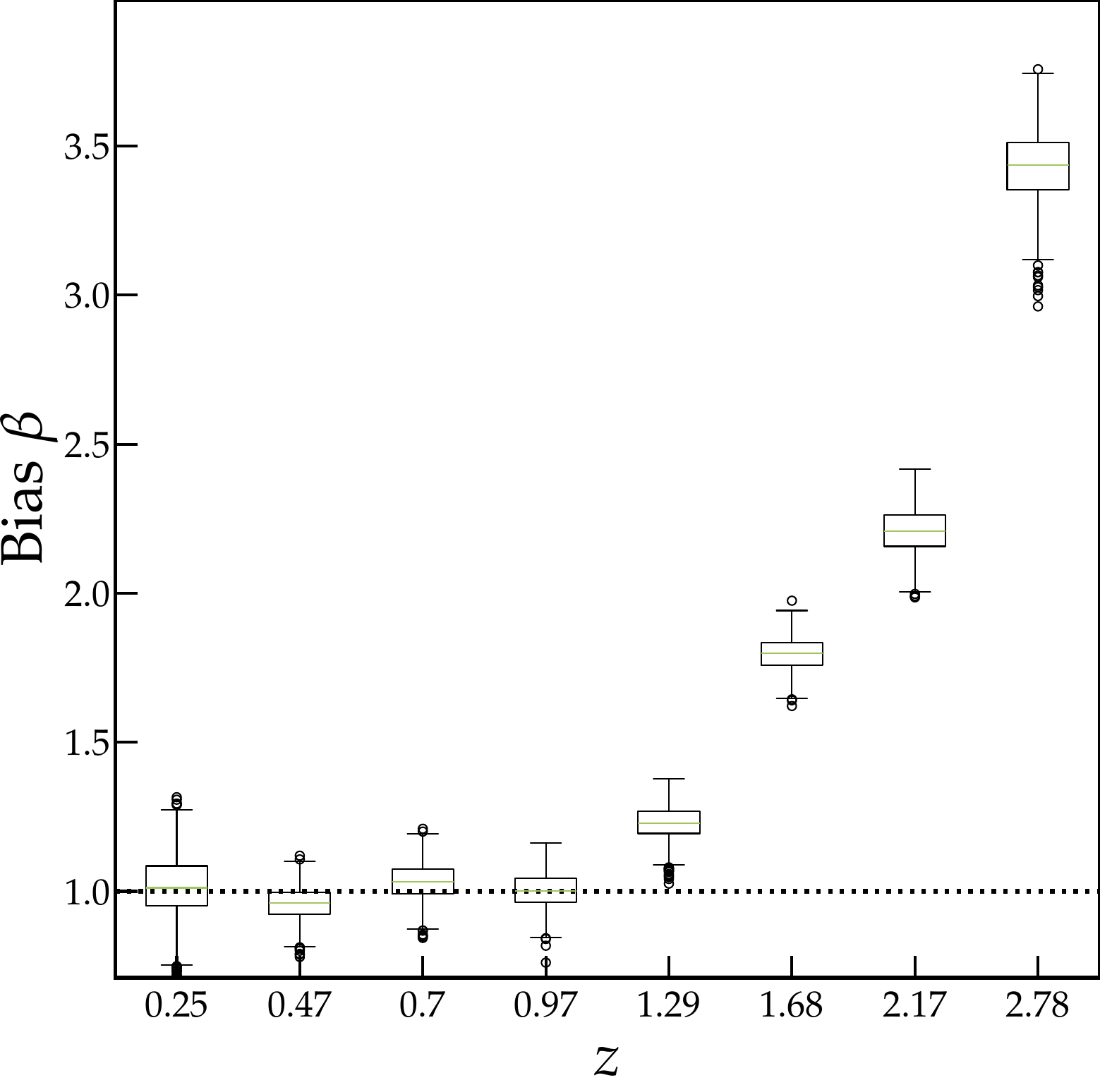}
    	\caption{The boxplot summarizes the posterior distribution of the power-law exponent $\beta$ for eight distinct galaxy populations, corresponding to different redshift bins. The clear monotonic increase in the median value of $\beta$ with redshift indicates a steepening of the galaxy bias relation, indicating that the galaxy-matter scaling becomes steeper with redshift. At higher redshift the posterior constraints broaden, likely because the galaxy sample becomes sparser and the increased shot noise limits the precision with which the parameter can be inferred.}
    	\label{fig:bias_vs_z}
\end{figure}

The distribution of the galaxy bias parameter, $\beta$ (defined in equation \ref{eq:bias_model}), across the different redshifts provides a measurement of the inferred scaling between galaxy density and the underlying dark matter field. The plot \ref{fig:bias_vs_z} summarises the marginalised posterior distributions of $\beta$ in each of the eight redshift bins considered, allowing a direct comparison of median values and associated uncertainties across tracers and redshifts.

The posterior samples are drawn from the converged segments of the MCMC chains. For each redshift bin, the ensemble of sampled $\beta$ values is represented as a boxplot, where the central line indicates the median, the box spans the interquartile range, and the whiskers show the full range of the data, with outliers displayed individually. This representation captures the central tendency, dispersion, and skewness of the parameter constraints without assuming Gaussianity.

The boxplot indicates a clear trend of increasing $\beta$ with redshift. The median value rises from ${\sim}1.0$ at low redshift to ${\sim}3.5$ at high redshift.


\section{Systematic Contamination in \textit{Quaia Deep}  ($G<20.5$)}\label{Appx:SystContamination}

\begin{figure*}
    \centering
    \begin{subfigure}{0.48\textwidth}
        \centering
        \includegraphics[width=\textwidth]{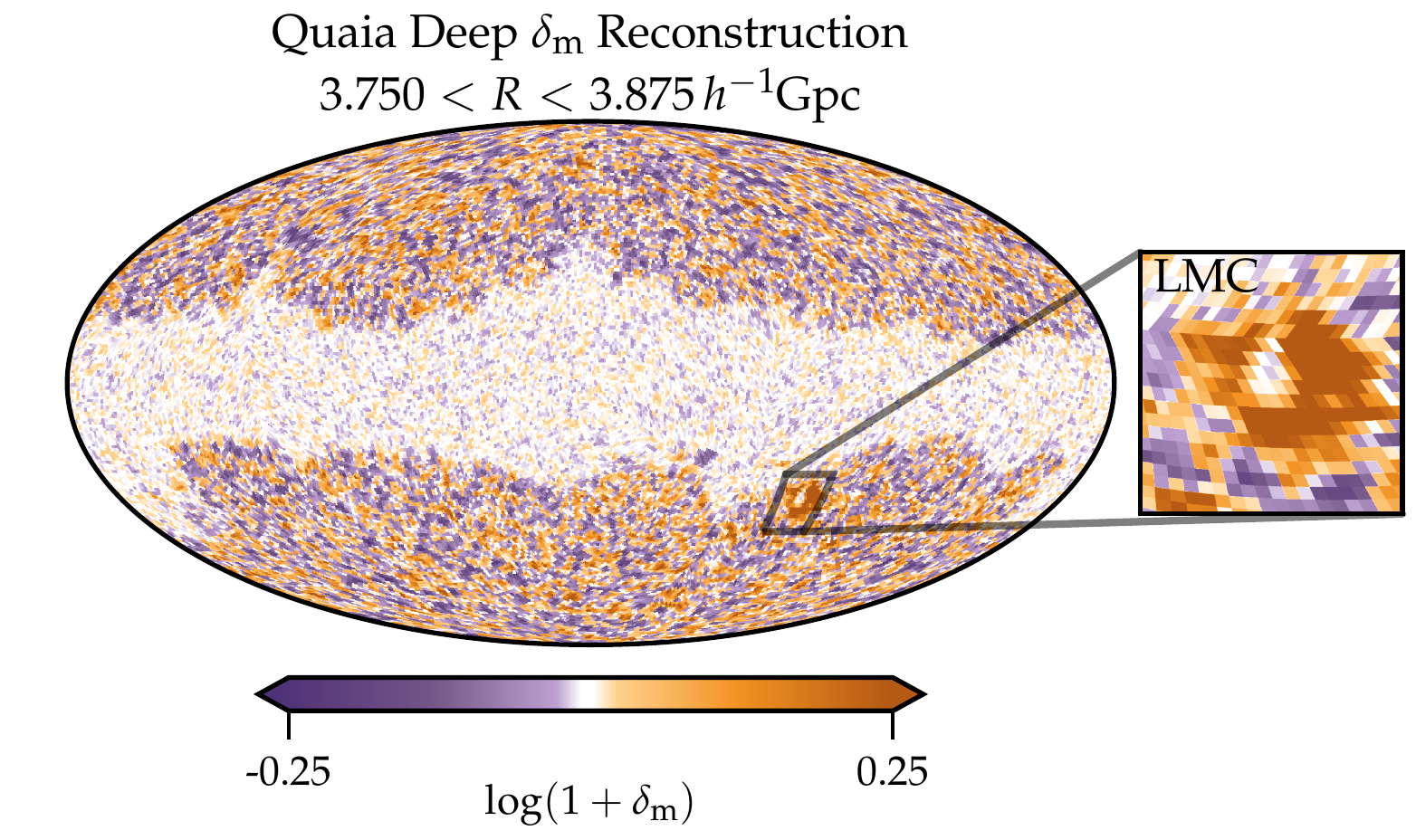}
        \caption{\textit{Quaia Deep} \borg{} Reconstruction.}
        \label{fig:QuaiaDeepBeast}
    \end{subfigure}
    \hfill
    \begin{subfigure}{0.48\textwidth}
        \centering
        \includegraphics[width=\textwidth]{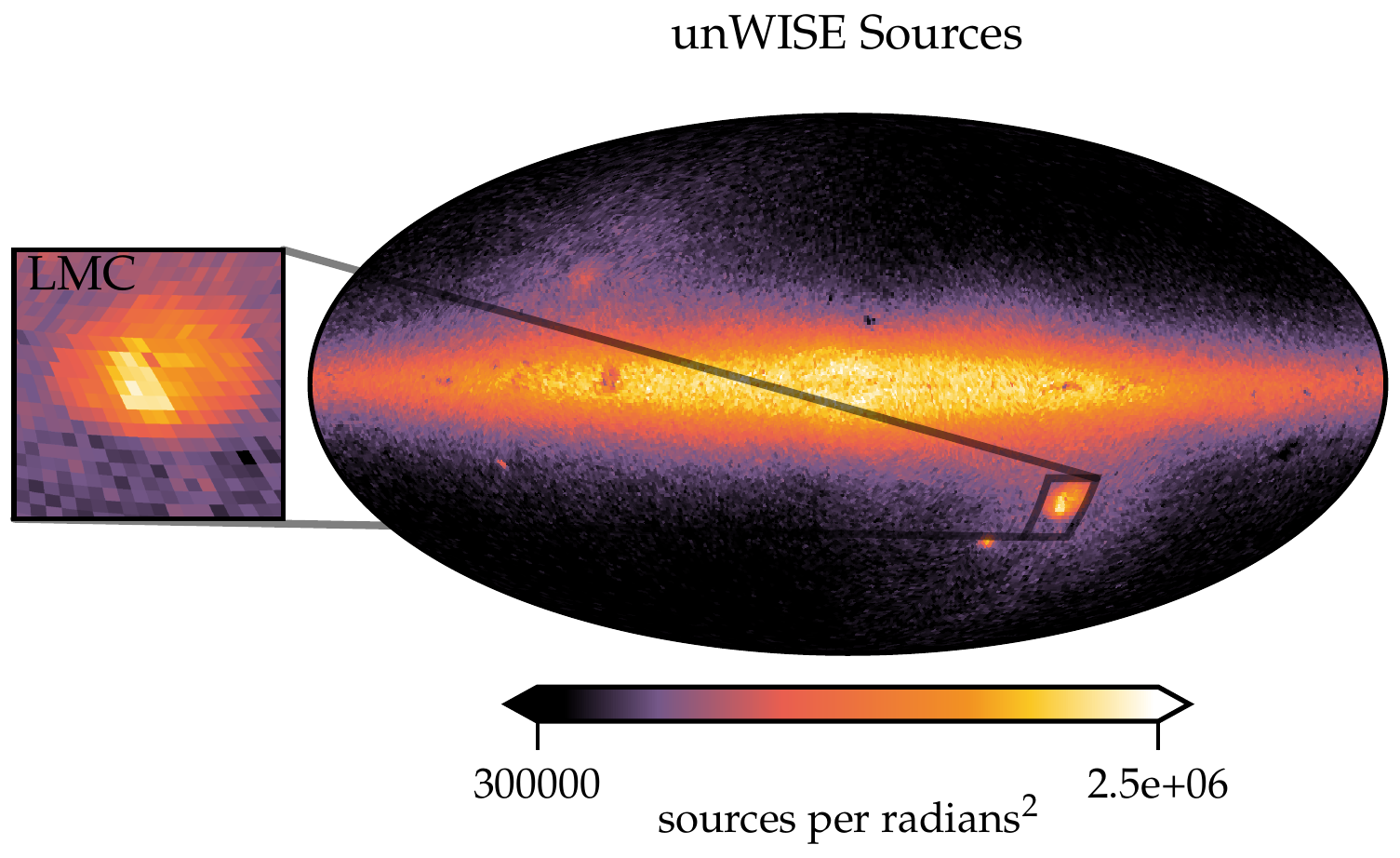}
        \caption{Foreground map of unWISE sources.}
        \label{fig:unWiseSources}
    \end{subfigure}

    \caption{Comparison between \textit{(a)} the reconstructed density field from the \textit{Quaia Deep} sample using the original angular selection function from SF24, and \textit{(b)} the density of unWISE sources used to construct the angular mask. Both panels focus on the region around the Large Magellanic Cloud (LMC). The prominent overdensity visible in panel (a) appears as a large, coherent ring-like structure in the reconstruction, extending over several Mpc. However, its close spatial correlation with the known foreground in panel (b) strongly suggests that this is not a physical feature, but rather the result of stellar contamination misclassified as high-redshift quasars in the range $20 < G < 20.5$. To mitigate this, we removed the affected region from the angular mask in our fiducial \textit{Quaia Deep Cut} analysis, as described in Section~\ref{sec:data}.}  
    \label{fig:thebeast}
\end{figure*}

In this appendix, we describe the motivations behind the sky cuts applied to the sample we refer to in Section~\ref{sec:data} as the \textit{Quaia Deep Cut}. These cuts were implemented to mitigate systematic contamination identified in the original \textit{Quaia} Deep sample provided by SF24, which includes all sources classified as quasars with $G<20.5$. We initially analysed both the \textit{Quaia Clean} ($G<20.0$) and \textit{Quaia} Deep ($G<20.5$) samples, using the catalogues and angular selection functions as released by  SF24, and performing the full Bayesian hierarchical inference with \borg{} at both $N=128$ and $256$.

In our preliminary tests, we did not apply any additional redshift or quality cuts and used the original angular selection mask. Upon analysis the original samples with \borg{}, we inspected the reconstructed three-dimensional matter density field and noticed a large and unphysical structure appearing at comoving distances between 3.75 and 3.875 $h^{-1}\mathrm{Gpc}$, i.e. at highest redshifts. This structure was spatially coherent over several tens of Mpc and located near the Galactic plane. It exhibited a signal-to-noise ratio substantially larger than any other structure in the reconstruction's samples, raising immediate concerns about its physical validity.

Visual inspection revealed that this spurious structure was co-spatial and of similar angular extent to the source counts in the Large Magellanic Cloud (LMC) region, as traced by unWISE sources. This strongly suggested that the feature was due to stellar contamination misclassified as high-redshift quasars rather than a real overdensity in the quasar distribution. Figure~\ref{fig:thebeast} shows the high-redshift reconstructed density field where the spurious structure is visible (Figure~\ref{fig:QuaiaDeepBeast}), alongside a map of unWISE sources highlighting the contaminated region (Figure~\ref{fig:unWiseSources}). The visual correlation between the two and the fact that such massive structure is not present in our \textit{Quaia Clean} reconstructions clearly indicates the non-cosmological origin of the feature. 

To mitigate this issue, we introduced additional sky cuts targeting regions of known contamination, particularly around the LMC. These cuts are described in Section~\ref{sec:data} and were implemented directly in the angular mask, shown in Figure~\ref{fig:QuaiaDeepCutCompletness}. We then re-ran all subsequent \textit{Quaia Deep} analyses using this conservative angular selection function. The total area removed from the original mask due to this contamination amounts to $28114.68 \rm{deg}^2$, only 4.2\% smaller than the original angular selection function by SF24. We note also that the latest cosmological results by the \textit{Quaia} Team, focused on Primordial Non-Gaussianities, also removes the SMC and \& LMC from their selection function, as shown in figure 2 of \cite{2025arXiv250420992F}.

\section{Technical Details for Method}
\label{Appx:TechDets}

This appendix provides an overview of the computational components underlying our inference pipeline. We begin by outlining the preparation of the three-dimensional survey window, including numerical integration, convergence control, and masking refinements. We then summarise the Hamiltonian Monte Carlo scheme used to sample the high-dimensional posterior, highlighting its formulation, integration strategy, and mass-matrix choice. Subsequently, we describe the likelihood and bias model that connects the evolved matter density to galaxy counts through a local, power-law prescription combined with Poisson statistics. Finally, we detail the initialisation of the chains and the convergence diagnostics employed to ensure robustness of the posterior samples. These sections detail the methodological foundations of our analysis and motivate the assumptions, numerical procedures, and validation steps that underpin the results.

\subsection{Additional Data Preparation}

The three-dimensional survey window is constructed by evaluating the selection function on the voxel grid used in the reconstruction. For each voxel, \borg{} computes the mean completeness by integrating the angular and radial selection functions over the voxel volume with a Monte-Carlo Miser integrator. The procedure begins with an angular-only integration, providing an estimate of the sky completeness averaged over the voxel. Voxels with angular completeness below a fixed threshold (0.5 in the implementation) are discarded at this stage. For voxels that pass this pre-check, the full selection function -- combining sky completeness and radial selection -- is integrated to a prescribed relative precision, and the resulting value is normalised by the voxel volume.

The numerical accuracy of each integral is adaptively controlled. Both the angular-only and full three-dimensional integrations are repeated with increasing numbers of Monte-Carlo calls until the estimated relative error falls below the specified tolerance or a maximum iteration count is reached. Voxels for which the integrator fails to converge are conservatively assigned zero completeness. The relative-error estimates are stored internally, allowing inspection of convergence quality when needed. All computations are parallelised with OpenMP and executed across MPI domains, ensuring scalability for large grids.

After the primary evaluation, the mask is further refined. Voxels flagged as insufficiently covered in the angular pre-check are explicitly set to zero in the final window field, and the grid is trimmed by removing a configurable number of boundary layers to restrict the analysis to well-sampled interior regions. The resulting three-dimensional completeness field forms the effective survey window used by the forward model and the likelihood.

\subsection{Hamiltonian Monte Carlo sampler}

Efficient exploration of the high-dimensional LPT--Poissonian posterior requires a sampling strategy capable of circumventing the limitations of random-walk Metropolis--Hastings algorithms. In standard schemes, the acceptance probability deteriorates rapidly with dimensionality, rendering them numerically impractical for inference problems involving millions of parameters. Hamiltonian Monte Carlo (HMC) mitigates this issue by introducing auxiliary momenta and interpreting the negative log-posterior as a potential energy,
\begin{equation}
    \psi(\boldsymbol{x}) = -\ln P(\boldsymbol{x}),
\end{equation}
thereby enabling the construction of a Hamiltonian system whose deterministic trajectories follow gradients of the posterior. This approach suppresses random-walk behaviour and maintains high sampling efficiency even in strongly non-linear regimes.

The joint distribution of parameters $\boldsymbol{x}$ and momenta $\boldsymbol{p}$ is defined through the Hamiltonian
\begin{equation}
    H(\boldsymbol{x},\boldsymbol{p}) = \frac{1}{2}\,\boldsymbol{p}^{\mathsf{T}} M^{-1} \boldsymbol{p} + \psi(\boldsymbol{x}),
\end{equation}
where $M$ is a tunable mass matrix. By construction,
\begin{equation}
    \exp[-H(\boldsymbol{x},\boldsymbol{p})] = P(\boldsymbol{x})\,\mathcal{N}(\boldsymbol{p}\,|\,0,M),
\end{equation}
so that marginalizing over $\boldsymbol{p}$ recovers the target posterior. The sampler proceeds by first drawing momenta from the Gaussian kinetic term and then evolving $(\boldsymbol{x},\boldsymbol{p})$ along Hamilton’s equations,
\begin{align}
    \frac{d\boldsymbol{x}}{dt} &= M^{-1}\boldsymbol{p}, \\
    \frac{d\boldsymbol{p}}{dt} &= -\nabla \psi(\boldsymbol{x}),
\end{align}
for a pseudo-time interval~$\tau$. A Metropolis--Hastings accept-reject step ensures correctness. In the limit of exact integration, energy conservation implies unit acceptance probability, while in practice a symplectic leapfrog integrator preserves reversibility and maintains high acceptance rates.

For large-scale structure inference, the forces required in $\nabla\psi$ include both the Gaussian prior and the non-linear likelihood contribution arising from the forward model. Writing
\begin{equation}
    \psi(\delta_{\mathrm{IC}}) = \psi_{\mathrm{prior}}(\delta_{\mathrm{IC}}) + \psi_{\mathrm{like}}(\delta_{\mathrm{IC}}),
\end{equation}
the prior gradient is 
\begin{equation}
    \nabla\psi_{\mathrm{prior}} = S^{-1}\delta_{\mathrm{IC}},
\end{equation}
while the likelihood gradient is obtained through adjoint differentiation of the LPT mapping, redshift-space distortions, biasing, and Poisson sampling. These operations form a sequence of linear transformations acting on the density field and permit efficient evaluation of the Hamiltonian forces required for each leapfrog step. The leapfrog “kick-drift-kick’’ integrator is employed due to its symplectic nature, numerical stability, and exact reversibility, all of which are essential for maintaining detailed balance.

The efficiency of the sampler depends sensitively on the choice of the Hamiltonian mass matrix $M$. For computational tractability it must remain diagonal, and empirical evidence shows that choosing it proportional to the inverse cosmological power spectrum, $M = S^{-1}$, yields excellent performance. This choice reflects the variance structure of the density modes, accelerates exploration of the parameter space, and improves convergence. In combination with the symplectic integrator and gradient-informed proposals, HMC provides a rigorous and scalable framework for sampling from the non-linear LSS posterior, enabling reliable uncertainty quantification in complex cosmological inference tasks.

\subsection{Likelihood and Bias Model}
\label{sect:lh_and_bias}

The ``galaxy bias problem'' refers to the translation between the observed distribution of galaxies and the underlying dark matter density field predicted by gravitational-only models. While dark matter provides the gravitational framework for structure formation, galaxy formation is influenced by complex astrophysical processes, including feedback from star formation, mergers, and environmental interactions. As a result, galaxies do not trace the dark matter field in a simple proportional manner, but are instead considered to be \textit{biased} tracers. For a more extensive description, we refer to an extensive review on the matter by \cite{2018PhR...733....1D}.

In this work, we adopt a \textit{local} bias model, which accounts for the formation of galaxies in over-dense regions in the quasi-linear regime. This model establishes a statistical relationship between the evolved dark-matter density field and the galaxy distribution, enabling the algorithm to infer the large-scale structure while marginalising over uncertainties in galaxy formation physics.

We adopt a power-law bias model to describe the relationship between the local number density of galaxies and the evolved dark-matter density field. This model assumes that the galaxy number density $n(z)$ is an exponential function of the matter overdensity $\delta_{\rm{m}}$, parametrised as \citep{2015JCAP...01..036J}
\begin{equation}
    n_p^{\mathrm{pred}, c}(z) = \bar{n}^c \left [ 1 + \delta_{\mathrm{m}, p}(z) \right ]^{\beta^c},
    \label{eq:bias_model}
\end{equation}
where $\bar{n}$ is the mean galaxy number density, $\delta_{\rm m}$ is the matter overdensity field, $\beta$ is the bias exponent, and the index $p$ and $c$ stand for the voxels and different catalogues, respectively. Both $\bar{n}^c$ and $\beta^c$ are free parameters in the model and are allowed to vary within values above zero.

Given a predicted biased galaxy field, we model the observed galaxy counts using a Poisson likelihood, accounting for the discrete nature of galaxy observations. The likelihood is formulated as:
\begin{align}
    \mathcal{P}\left(n^{\rm obs}(z) \,\middle|\, n^{\rm pred}(z)\right) = \prod_p^{N_{\rm vox}} \frac{\left[ \mathcal{R}_{ p}n_p^{\text{pred}}(z)\right]^{n_p^{\text{obs}}(z)}}{n_p^{\text{obs}}(z)!} \, e^{- \mathcal{R}_{p}n_p^{\text{pred}}(z)} \, .
\end{align}

\subsection{Initialisation of Chains and Convergence Test}

In order to start far from the target posterior, the MCMC chains are initialised with initial conditions drawn from a Gaussian distribution with one-tenth the expected mean. The bias parameter $\beta$ was initialised at a value of 2.0 for all the catalogues. To prevent the bias parameters from compensating for the initially low amplitude of the initial conditions, the bias parameters were held fixed until the amplitude of the initial conditions reached approximately 70\% of the expected value, which typically occurred after the first $\sim$1,000 samples. After excluding the first 5,000 burn-in samples of the chain, our fiducial run retains approximately 100,000 MCMC samples for the final analysis.

We performed standard convergence diagnostics to assess the reliability of the posterior samples. Specifically, we computed the Gelman-Rubin $R$ statistic and the Effective Sample Size (ESS) for representative summary statistics, field moments, and bias parameters. Split into subchains, the chains achieved $R - 1 < 0.05$, with ESS values in the hundreds for the bias parameters. We further monitored the correlation length of Fourier modes to verify stationarity. A more detailed account of the convergence checks, including diagnostic plots and the quantitative results of these tests, is provided in Appendix~\ref{Appx:convergence}.

\section{MCMC Diagnostics \& Convergence}
\label{Appx:convergence}

\subsection{Gelman-Rubin test}
\begin{figure}
	\centering
        \includegraphics[width=\columnwidth]{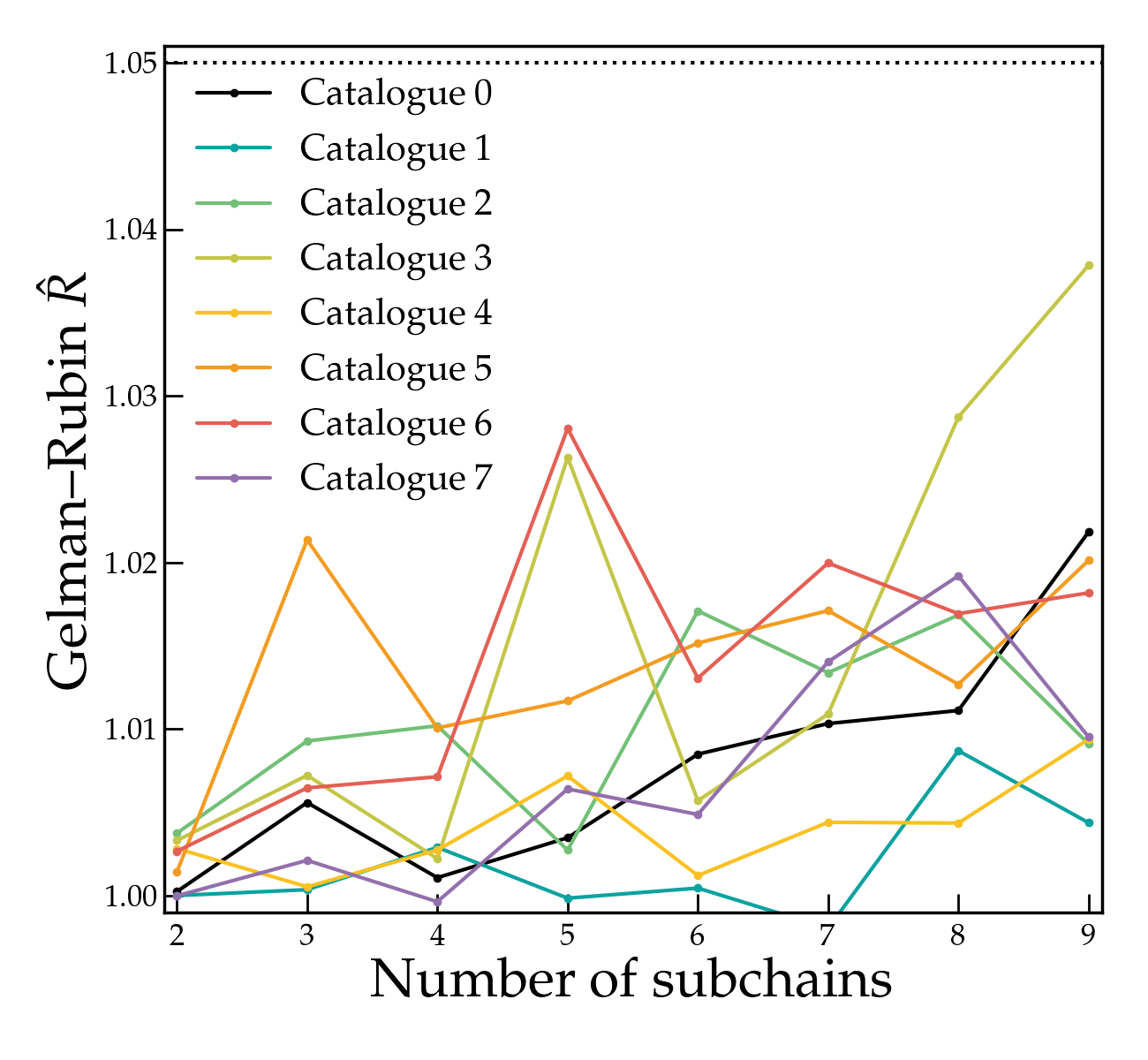}
	\caption{Gelman--Rubin convergence diagnostic $\hat{R}$ \citep{gr_test_1992} for the $\beta^c$ bias parameter across the eight catalogues for the high-resolution \textit{Quaia Deep Cut} analysis. The statistic is evaluated by artificially splitting each MCMC chain into $N_{\mathrm{sub}}=2,\dots,9$ subchains, discarding $1500$ samples between consecutive segments. Values of $\hat{R}$ close to unity indicate good mixing, with the dotted line marking the commonly used threshold $\hat{R}=1.05$. The results show that all catalogues remain consistent with convergence under varying subchain splits, with deviations at the few-percent level at most.}
	\label{fig:gr_test}
\end{figure}
To assess the convergence of our MCMC chains, we employ the Gelman-Rubin diagnostic \citep{gr_test_1992}, which compares the variance between multiple chains to the variance within each chain. A Gelman-Rubin statistic close to unity indicates that the chains have converged to the same target distribution, whereas values significantly greater than one suggest that the chains are still exploring different regions of parameter space. This diagnostic is particularly useful for identifying parameters with slow convergence.

In our analysis, we compute the Gelman-Rubin statistic for all power-law bias parameters ($\beta^c$, in equation~\ref{eq:bias_model}). Since we only have a single chain for each catalogue, we artificially split it into a varying number of subchains, with the subchains separated by 1500 discarded samples to reduce correlations between segments. The risk of underestimating the variance due to autocorrelation is thus mitigated by this gap, ensuring that each subchain provides an approximately independent estimate of the posterior variance.

We show the results of the test in Figure \ref{fig:gr_test}. Across all choices of the number of subchains, the statistics are less than 1.05, demonstrating convergence. These results confirm that the posterior distributions of the inferred quantities are sampled well, and suitable for subsequent statistical analysis and comparison with theoretical predictions.

\subsection{Correlation Length of Power Spectrum Modes}

To assess the convergence of our Markov chains, we analyse the correlation length of the inferred power spectrum modes. The correlation length quantifies the number of iterations required for the samples of a given mode to become effectively uncorrelated, providing a complementary measure to traditional diagnostics such as the Gelman-Rubin statistic.

\begin{figure}
	\centering
        \includegraphics[width=\columnwidth]{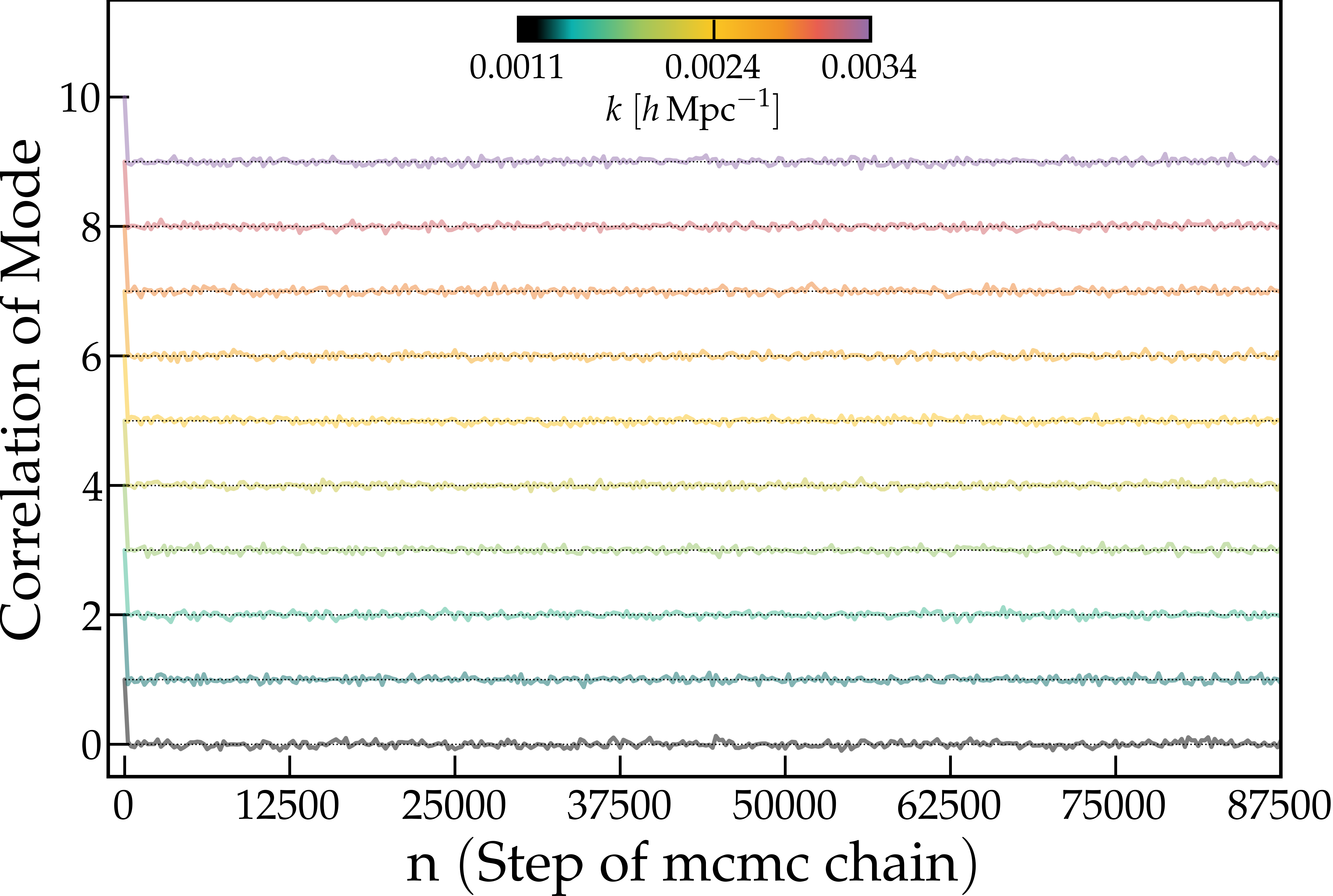}
	\caption{Autocorrelation of selected power spectrum modes. The plot shows the normalised autocorrelation function of a subset of power spectrum modes sampled from the MCMC chain of the \textit{Quaia Deep Cut} analysis. Each curve represents a single mode, offset vertically for clarity, and color-coded according to the mode's wavenumber $k$ (see colorbar). Only modes within the range $0.0011 \le k \le 0.0034\ h\,\mathrm{Mpc}^{-1}$ are shown. However, similar results are found for modes not included here. The colorbar above the figure provides the mapping between color and the wavenumber $k$ in units of $h\,\mathrm{Mpc}^{-1}$. This visualisation demonstrates the short correlation length of the chain across a wide range of modes, and highlights the range of independent sampling in each mode.}

	\label{fig:pk_corr_length}
\end{figure}

In Figure~\ref{fig:pk_corr_length}, we present the estimated correlation lengths for all Fourier modes of the inferred density fields. The correlation lengths are generally short relative to the total chain length, indicating that our ensemble has effectively explored the posterior distribution and that the samples can be treated as approximately independent for statistical analysis. Accordingly, the effective sample sizes of individual modes lie in the range of $10^3$-$10^4$, further demonstrating the efficiency of the sampling procedure and the reliability of the inference.

\subsection{Cross-studies between runs/set-ups}

To assess the robustness and convergence of our inference, we compare results from the high-resolution analyses of \textit{Quaia Deep Cut} and \textit{Quaia Clean}. By performing this comparison, we quantify the agreement between the reconstructed mean density fields using metrics such as the overall shape and amplitude of the present-day dark-matter distribution. This systematic evaluation provides a test of the reliability of our method and its ability to infer cosmological structures consistently across similar datasets.

\begin{figure}
	\centering
        \includegraphics[width=\columnwidth]{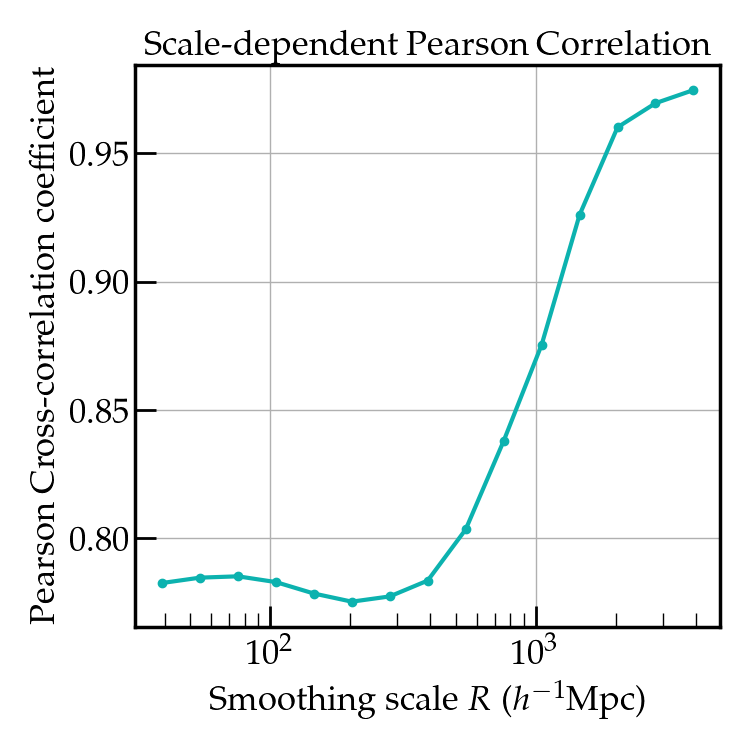}
	\caption{Scale-dependent Pearson cross-correlation coefficient between the mean density fields reconstructed from the \textit{Quaia Clean} and \textit{Quaia Deep Cut} chains. The correlation is evaluated within the survey mask, ensuring that only the observed volume contributes to the statistic. The correlation remains high across a broad range of smoothing scales, demonstrating the statistical consistency and robustness of the inferred structures. The persistence of the correlation as a function of scale indicates that the recovered density field is not sensitive to the specific realisation, reinforcing that we are capturing the large-scale features correctly.}
	\label{fig:pearson_ccc}
\end{figure}

Figure \ref{fig:pearson_ccc} presents the scale-dependent Pearson cross-correlation coefficient between the two reconstructions, computed within the survey mask to account only for the observed volume. The correlation remains high over a wide range of smoothing scales, demonstrating that the large-scale features of the inferred density field are consistently recovered. The persistence of the correlation across scales indicates that the analysis reliably captures the large-scale structure of the cosmic density field.

\section{Additional Validation of the Inferred Velocity Field}
\label{Appx:velocity_dipole}

\begin{figure*}
    	\centering
            \includegraphics[width=2\columnwidth]{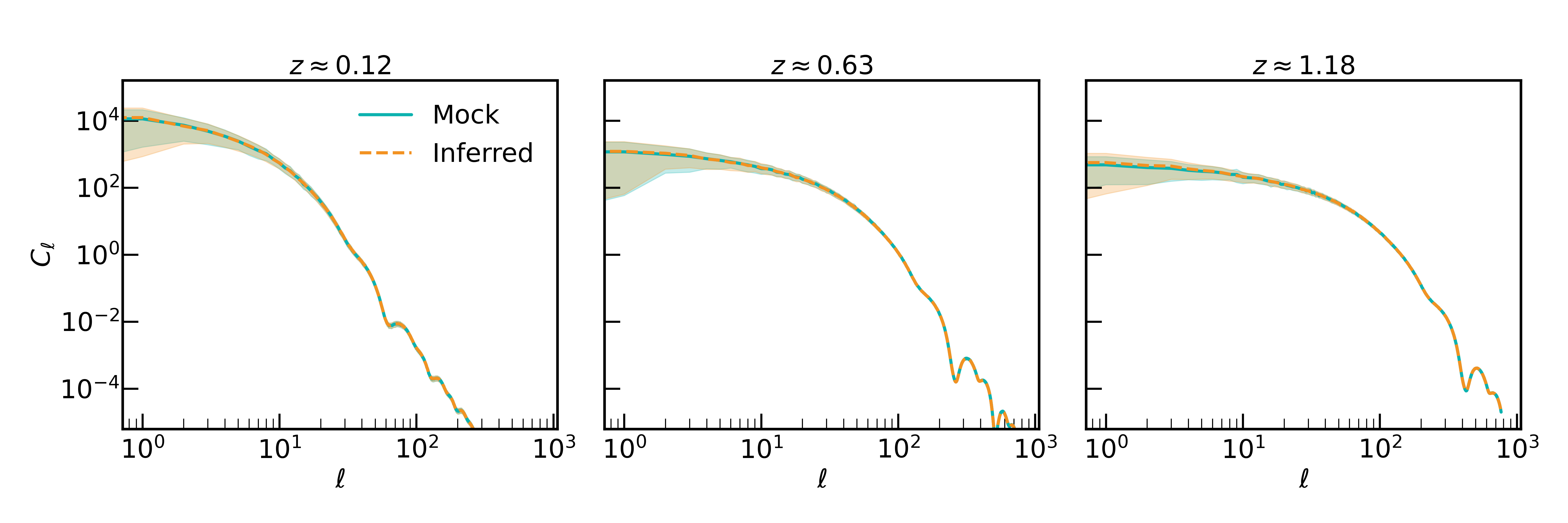}
    	\caption{Comparison of angular power spectra $C_\ell$ for three representative redshift bins.
            Solid blue lines show the mean $C_\ell$ from the mock velocity fields, 
            while the shaded regions indicate the $\pm 1\sigma$ standard deviation across the ensemble. 
            Dashed red lines correspond to the mean $C_\ell$ from the data-constrained velocity fields, 
            with shaded regions representing the $\pm 1\sigma$ uncertainties. 
            Panels from left to right correspond to increasing redshift values. 
            The comparison demonstrates that the inferred velocity fields reproduce the statistical properties 
            of the mock fields across the range of angular scales and redshifts considered.}
    	\label{fig:angular_vel_check}
\end{figure*}

\begin{figure*}
    	\centering
            \includegraphics[width=2\columnwidth]{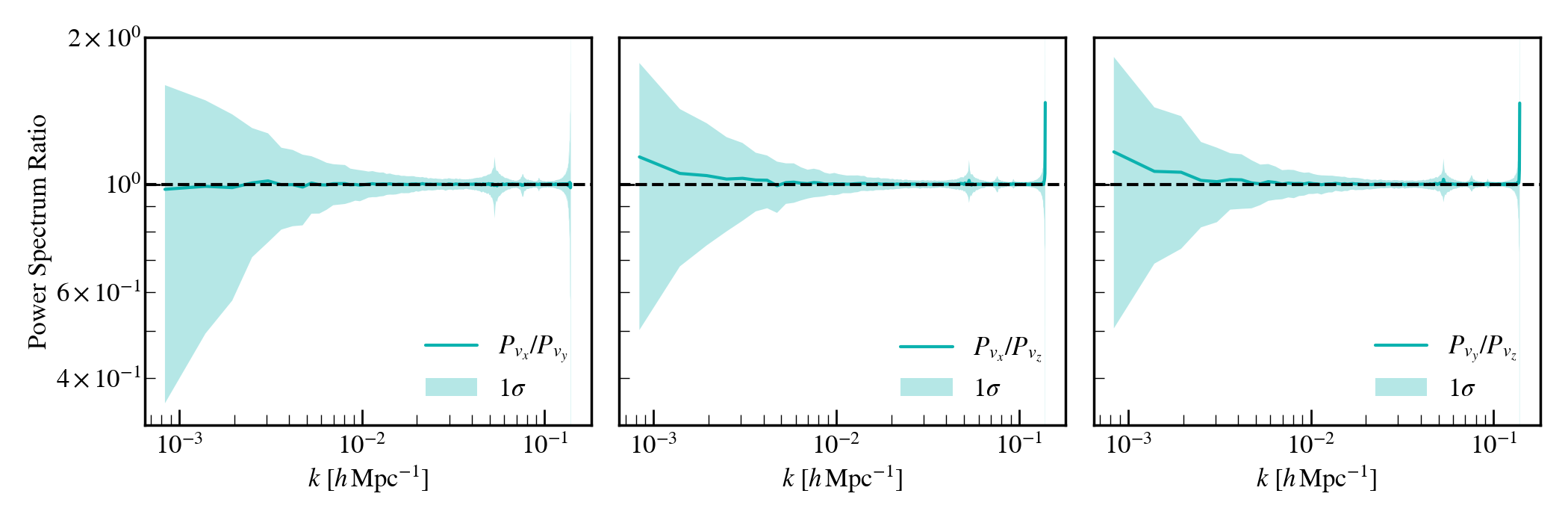}
    	\caption{Comparison of the ratios of the three-dimensional velocity power spectra for the Cartesian components. The three panels show the mean ratios $P_{v_x}/P_{v_y}$, $P_{v_x}/P_{v_z}$, and $P_{v_y}/P_{v_z}$, with shaded regions indicating the corresponding $\pm 1\sigma$ scatter across the ensemble. The ratios remain close to unity over the full range of wavenumbers, demonstrating that the reconstructed velocity field exhibits no significant directional bias and is consistent with statistical isotropy.}       
    	\label{fig:pk_vel_check}
\end{figure*}

The main goal of this analysis is to assess the fidelity of the inferred velocity field and to evaluate its consistency with a mock dataset based on the standard $\Lambda$CDM cosmology. By comparing the inferred and mock velocity fields, we aim to validate the performance of the reconstruction framework and identify any potential systematic biases or scale-dependent deviations. In particular, we focus on quantifying the level of agreement in both amplitude and shape of the velocity power spectra and angular statistics across spatial scales and Cartesian components. This comparison serves as a stringent test of the velocity field reconstruction and provides insight into the reliability of the recovered large-scale velocity field.

First, we generated random velocity fields based on the standard $\Lambda$CDM cosmology, using random realisations of the initial conditions. 
These mock velocity fields were evolved with the same forward model employed for the data-constrained velocity reconstructions. Next, we computed the angular power spectra for a range of redshift bins with a uniform width of $\Delta z = 0.05$. These spectra were evaluated for both the data-constrained and mock velocity fields, allowing a direct comparison of their large-scale angular statistics. In addition, we computed the three-dimensional velocity power spectra for the full simulation cube, separately for the $x$-, $y$-, and $z$-components of the velocity field, again for both the inferred and mock datasets. In addition, we computed the velocity power spectra for each of the eight subcubes (with box size $L/2$), in order to probe potential spatial variations and assess the local consistency of the reconstructed velocity field. For each case, the analysis was extended across the full ensemble of available samples, enabling a statistical comparison between the two datasets in terms of both the mean and the variance. Figure~\ref{fig:angular_vel_check} presents the mean and $\pm1\sigma$ dispersion of the angular power spectra $C_\ell$ for each redshift bin. Figure~\ref{fig:pk_vel_check} presents the ratios of the three-dimensional velocity power spectra for the different Cartesian components. The three panels show the scale dependence of $P_{v_x}/P_{v_y}$, $P_{v_x}/P_{v_z}$, and $P_{v_y}/P_{v_z}$, providing a direct test of the directional consistency of the reconstructed velocity field. Across all wavenumbers, the ratios remain close to unity within their statistical uncertainties, indicating that the reconstruction does not introduce anisotropies and that the inferred velocity field is statistically isotropic over the scales probed.

The comparison between the inferred and mock datasets reveals overall agreement across most scales. Both the angular power spectra $C_\ell$ and the three-dimensional velocity power spectra $P(k)$ exhibit consistent amplitudes and shapes. The agreement in normalisation between the mock and inferred power spectra further confirms that the reconstruction preserves the expected amplitude of large-scale modes. The component-wise and subcube analyses show no significant directional dependence, demonstrating that the recovered velocity field is statistically isotropic within the uncertainties. These diagnostics provide sufficient evidence that the inference framework successfully reproduces the statistical characteristics of the $\Lambda$CDM mock dataset, both globally and locally, thereby validating the reliability of the reconstructed velocity field.

\section{Modelling of the CMB Lensing Cross-Correlations Signal}\label{Appx:cmb_signal_modelling}

To evaluate whether the signal obtained from cross-correlating the $\kappa_{\rm BORG}$ map with the CMB lensing convergence map, $\kappa_{\rm CMB}$, is consistent with theoretical expectations, we model the signal using the Core Cosmology Library (\texttt{CCL}, \citealt{2019ApJS..242....2C}). The theoretical angular power spectra between two different tracers, $a$ and $b$, in \texttt{CCL} is defined as
\begin{equation}
    C_{\ell}^{a,b} = 4\pi \int P_{\Phi}(k)\Delta_{\ell}^{a}(k)\Delta_{\ell}^{b}(k)\frac{\text{d}k}{k},
    \label{eq:CCL_Cls}
\end{equation}
where $P_{\Phi}(k)$ is the dimensionless primordial power spectrum of curvature perturbations, and $\Delta_{\ell}^{a}(k)$ and $\Delta_{\ell}^{b}(k)$ are the window (or transfer) functions for the respective tracers.

In our case, the two tracers are $\kappa_{\rm BORG}$ and $\kappa_{\rm CMB}$. For the CMB lensing convergence map from \textit{Planck}, the window function is given by \citep{2000PhRvD..62d3007H,2019ApJS..242....2C}:
\begin{equation}
    \Delta^{\kappa_{\rm CMB}}_{\ell}(k) = - \frac{\ell (\ell + 1)}{2} \int_{0}^{z_{\rm CMB}} \frac{T_{\phi+\psi}(k,z)}{H(z)} \left(\frac{\chi_{\rm CMB} - \chi(z)}{\chi(z)\chi_{\rm CMB}}\right) \text{d}z,
\end{equation}
where $z_{\rm CMB}$ is the redshift of the last scattering surface, $T_{\phi+\psi}(k,z)$ is the \texttt{CAMB} transfer function \citep{Lewis:1999bs, Howlett:2012mh} for the Newtonian-gauge scalar metric perturbations (see \citealt{2019ApJS..242....2C}, section 2.4.1 for more details), and $H(z)$ is the Hubble expansion rate at a given redshift $z$.

For the \borg{} samples, the window function corresponds to the CMB convergence of a ``slab of matter'' convolved with the sample's radial response, $\mathcal{R}(z)$:
\begin{equation}
    \Delta^{\kappa_{\rm BORG}}_{\ell}(k) = - \frac{\ell (\ell + 1)}{2} \int_{z_{\rm min}}^{z_{\rm max}} \frac{T_{\phi+\psi}(k,z)}{H(z)} \left(\frac{\chi_{\rm CMB} - \chi(z)}{\chi(z)\chi_{\rm CMB}}\right)\mathcal{R}(z) \text{d}z,
    \label{eq:CMB_Lensing_kernel}
\end{equation}
since \borg{} $\delta_{\rm m}$ samples trace the matter density field from $z_{\rm min} = 0.174$ to $z_{\rm max}=3.0$ as determined by the radial selection function shown in the shaded region of Fig \ref{fig:20_5_rsf}.

This approach ensures that the modelled signal captures the physical correlations expected from the interaction between the reconstructed matter density field ($\kappa_{\rm BORG}$) and the CMB lensing convergence ($\kappa_{\rm CMB}$). However, owing to the complexity of the \borg{} forward model, this theoretical description necessarily represents an idealised approximation to the expected cross-correlation signal. In particular, while the radial selection function of the \textit{Quaia} sample is explicitly accounted for, the effective biasing and transfer functions introduced by the \borg{} inference framework are not fully captured within the \texttt{CCL} formalism.
As a result, the predicted $C_{\ell}^{\kappa_{\rm BORG}\kappa_{\rm CMB}}$\footnote{Abbreviated to $C_{\ell}^{\kappa_{\rm B}\kappa_{\rm C}}$ in the main text.} should be interpreted as a theoretically motivated best-case template for the cross-correlation
signal rather than a fully forward-modelled prediction. This motivates the introduction of a free amplitude parameter in Section~\ref{sec:amplitude-fitting}, which absorbs residual mismatches between the idealised template and the measured cross-correlations while preserving sensitivity to the large-scale signal encoded in their overall shape.
However, to enable a meaningful comparison between theory and measurement, the effects of voxel-to-pixel in the \borg{} data model must be accounted for explicitly, as described in the following section.

\subsection{The \textit{Voxel-to-Pixel} Window}
\begin{figure}
    	\centering
            \includegraphics[width=\columnwidth]{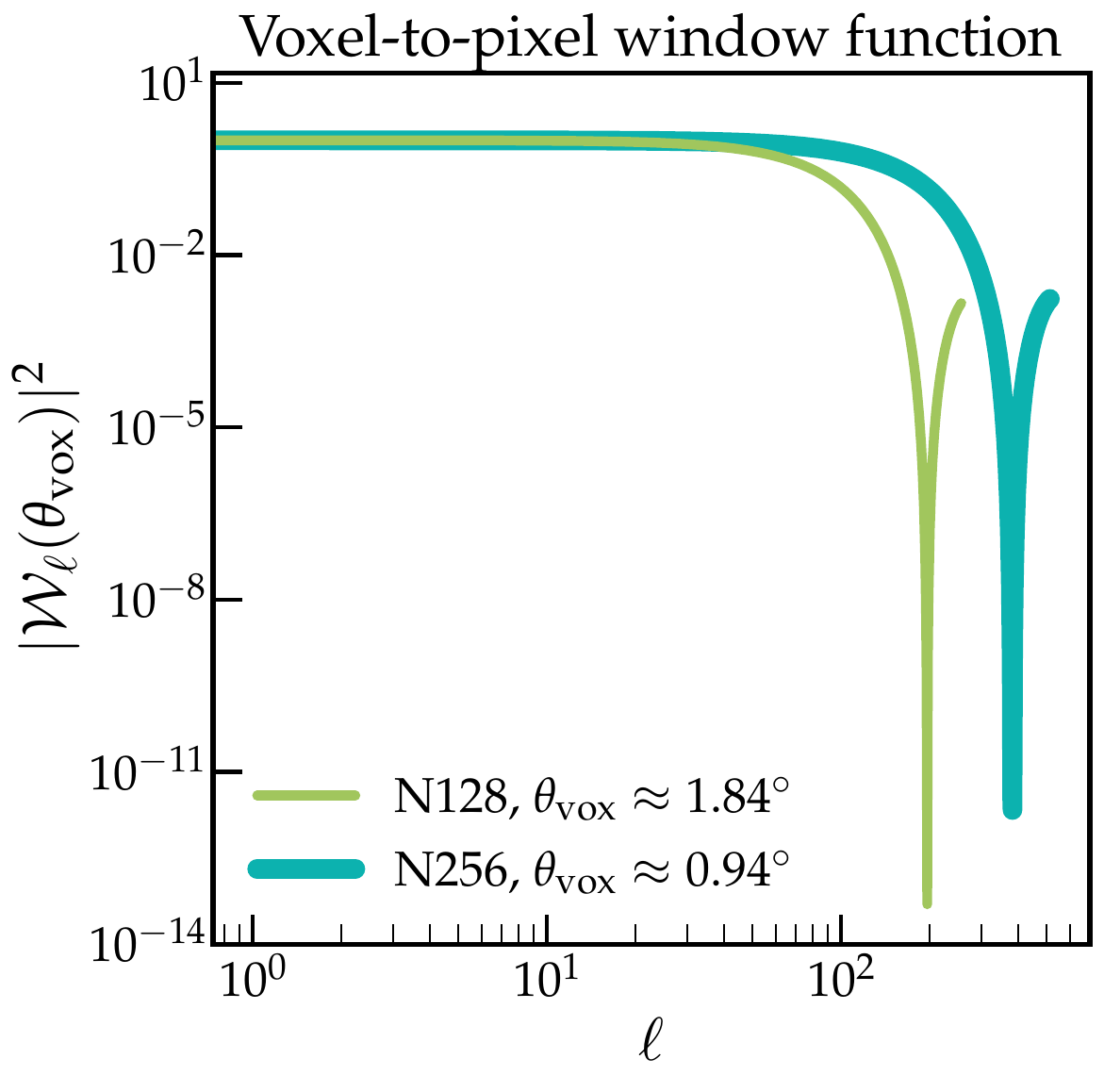}
    	\caption{The voxel-to-pixel window function for different \borg{} resolutions as described by Eq. \eqref{eq:voxpixwin}. Scales around $\theta^{}_{\rm VOX}\ell \gg 1 $ are suppressed by the window function, limiting our validation via CMB lensing cross-correlations to large scales.}
    	\label{fig:vox2pix_window}
\end{figure}
When projecting the 3D \borg{} matter density field onto the sky to compute the convergence, the finite resolution of the voxelised data must be accounted for. This process introduces an additional smoothing effect, encapsulated in the \textit{voxel-to-pixel} window function. While the CIC kernel has been deconvolved from the \borg{} $\delta_{\rm m}(\mathbf{r})$ samples, the projection of voxels onto spherical pixels imposes an angular resolution limit that affects the angular power spectrum.

The angular size of a voxel projected onto the sky is defined as $\theta_{\rm vox} = \Delta L / R_{\rm max}$, where $\Delta L$ is the side length of a voxel and $R_{\rm max}$ is the maximum comoving distance to the observer within the box of size $L$. Approximating $k \approx \ell / R_{\rm max}$ and $k_{\rm Ny} \approx \pi / \Delta L$, the \textit{voxel-to-pixel} window function can be expressed as:
\begin{equation}
    \mathcal{W}_{\ell}(\theta_{\rm vox}) = \left[ \text{sinc}\left(\frac{\ell \theta_{\rm vox}}{2\pi}\right) \right]^2,
    \label{eq:voxpixwin}
\end{equation}
where $\text{sinc}(x) = \sin(x) / x$ is the normalised sinc function. This window function suppresses power at scales smaller than the angular size of the projected voxels, with significant damping for modes where $\ell \theta_{\rm vox} \gg 1$ as shown in Figure \ref{fig:vox2pix_window}. This suppression is intrinsic to the resolution of the \borg{} data model.

To account for this effect, the theoretical angular power spectra, $C_{\ell}$ must be corrected for the \textit{voxel-to-pixel} window. The observed angular power spectrum, $C^{\rm obs}_{\ell}$, is then given by:
\begin{equation}
    C^{\rm obs}_{\ell} = |\mathcal{W}_{\ell}(\theta_{\rm vox})|^2 C_{\ell},
\end{equation}
where $C_{\ell}$ is the uncorrected angular power spectrum, as calculated by \texttt{CCL} with Eq.~\ref{eq:CCL_Cls}. This correction ensures that the theoretical expectations account for the finite resolution of the projected data.

The impact of the \textit{voxel-to-pixel} window function is particularly pronounced at high $\ell$ modes, where the angular scales correspond to smaller features than the voxel size. This effect is illustrated in Figures~\ref{fig:cmb_x_borg_compare} and~\ref{fig:cmb_x_borg_deep}, where the black lines highlight the suppression of power due to voxelisation. Applying this correction is essential for accurate comparisons between the modelled and observed angular power spectra, allowing us to validate the signal we obtain in our cross-correlations and ensuring we can trust the scales of our reconstruction where such signal is detected.

\bsp	
\label{lastpage}
\end{document}